\documentclass[journal]{IEEEtran}
\usepackage{cite}
\usepackage{amsmath,amssymb,amsfonts}
\usepackage{algorithm}
\usepackage{tabularray} 
\usepackage{multirow}  
\usepackage{comment}
\usepackage{booktabs}
\usepackage{algpseudocode}
\usepackage{amsmath} 
\usepackage{algorithm}
\usepackage{graphicx}
\usepackage{textcomp}

\begin{document}
\title{SONAR: A Structure-Consistent Neural Operator for Null-Space-Aware Sparse View CT Reconstruction}

\author{Song Ni, Haijun Yu, Haodong Li, Changsheng Fang, Shuyi Fan, Yixing Huang, Hengyong Yu~\IEEEmembership{Fellow,~IEEE} 
\thanks{ This work involved human subjects in its research. The authors confirm that all human subject research procedures and protocols are exempt from the Institutional Review Board at the University of Massachusetts Lowell (IRB\# 23-043). Song Ni and Haijun Yu contributed equally to this work. All correspondence should be addressed to Y. Huang or H. Yu (email:huangyx@pku.edu.cn, hengyong-yu@ieee.org). } 
\thanks{Song Ni, Haodong Li, Changsheng Fang, Shuyi Fan, and Hengyong Yu are with the Department of Electrical and Computer Engineering, University of Massachusetts Lowell, Lowell, MA 01854. }
\thanks{
Haijun Yu and Yixing Huang are with the Institute of Medical Technology, Peking University Health Science Center, Beijing, China. 
}
}

\maketitle
\thispagestyle{empty}

\begin{abstract}
Sparse-view computed tomography (CT) reduces radiation dose and
acquisition time but remains severely ill-posed because incomplete
projections poorly constrain null-space information.
Existing learning-based methods often estimate this information in
high-dimensional image space, conflate physical measurement errors with
prediction errors, and depend on fixed discretizations. We propose
SONAR, a Structure-Consistent Neural Operator for Null-Space-Aware
Reconstruction. Instead of recovering the full null-space component,
SONAR predicts a low-dimensional null-space-aware representation from
the acquired projections as pseudo-measurements. It separates
measurement and pseudo-measurement residuals, lifts them into the image
domain through physics operators, and applies independent neural
operators to constrain their structural effects, thereby accommodating
admissible errors while suppressing unsupported structures. To support
cross-discretization reconstruction, an anisotropic U-shaped neural
operator models the periodic angular and nonperiodic detector dimensions
using direction-dependent continuous supports, while image-domain neural
operators re-discretize continuous kernels on target grids. These
components form an optimization-inspired unrolled network. Experiments
on simulated AAPM and clinical MARS photon-counting CT data demonstrate
consistent improvements across seen and unseen view settings and unseen
image resolutions. On AAPM dataset, SONAR improves PSNR by 1.87~dB at 62 views
and by 7.63~dB under zero-shot transfer to a $512\times512$ grid over
the strongest competing methods. SONAR also achieves the best overall
performance in all clinical settings evaluated, demonstrating
accurate, structurally reliable, and discretization-robust sparse-view
CT reconstruction.
\end{abstract}

\begin{IEEEkeywords}
Computed Tomography, sparse-view reconstruction, null-space learning, neural operators, anisotropic DISCO, sinogram completion.
\end{IEEEkeywords}

\section{Introduction}

In computed tomography (CT), acquisition of sparse-view data reduces the number of projections over the full angular range, providing an effective way to lower radiation dose, shorten acquisition time, and improve imaging efficiency. 
Sparse-view scheme is particularly attractive for CT systems that employ pulsed exposures or discrete projection acquisition and allow flexible control of the number of views, for example, on-board cone-beam CT (CBCT) systems for radiation therapy systems~\cite{chang2026patient, zhao2025deep}, interventional and intraoperative C-arm CBCT systems~\cite{jeon2026utility}, industrial CT~\cite{ahmad2026compensating} or micro-CT systems~\cite{yin2026three}. Especially, for Medipix detector-based MARS photon-counting CT (PCCT) systems~\cite{panta2018first, li2025deep}, which feature relatively slow scanning and more flexible acquisition control, sparse-view can not only reduce dose and acquisition time but also alleviate the computational burden associated with multi-energy data processing.

However, sparse-view reconstruction is a severely ill-posed inverse problem, becasue sparse measurements cannot uniquely determine all image components~\cite{natterer2001mathematics}. 
The resulting ambiguity resides in the null and near-null spaces of the imaging operator: null-space components are invisible to the acquired projections, whereas near-null-space components generate only weak responses that noise, discretization errors, and geometric mismatch can readily obscure. 
Sparser angular sampling further enlarges these poorly constrained subspaces, making directional edges, small structures, and high-frequency details increasingly difficult to reliably recover.

Over the past decade, deep learning has substantially advanced sparse-view CT reconstruction~\cite{wang2023review,li2026cross}. Despite these gains, two fundamental challenges remain: (1) recovering image information that lies in the null or near-null space of the imaging operator; (2) maintaining reconstruction performance across varying sampling densities and image discretizations.

\subsection{Null-Space Learning}

From the perspective of the imaging operator, most deep-learning-based methods for sparse-view reconstruction can be interpreted as estimating invisible or weakly visible information constrained by the acquired measurements~\cite{schwab2019deep, chen2020deep}.  
Depending on whether they explicitly construct and exploit a null-space projector, these methods can be broadly divided into explicit null-space learning and implicit null-space estimation. 

\subsubsection{Explicit Null-Space Learning}

Explicit null-space learning combines a null-space projector derived from the imaging operator with a neural network that primarily estimates information in the null space. 
Representative approaches include Deep Null Space Learning~\cite{schwab2019deep}, which projects the learned correction onto the null space to recover invisible components while preserving data consistency.
Deep Decomposition Learning Network (DDN)~\cite{chen2020deep} further separates range-space errors and null-space corrections to improve robustness to noise and model mismatch. 
More recently, Non-linear Projections of the Null-space (NPN)~\cite{jacome2026npn} replaces predicting full null-space information with estimating low-dimensional sub-null-space coordinates to reduce the complexity of learned null-space estimation. 

The strong generative capability of diffusion models further enables explicit generation of null-space components from a learned prior data distribution~\cite{song2021solving, yang2025ct, cheng2023null}. 
Representative approaches include Denoising Diffusion Null-Space Model (DDNM)~\cite{wang2022zero}, which preserves measurement-consistent components while updating primarily the null-space content during reverse diffusion. 
Conditional Score-based Null-space (CSN)~\cite{zhang2025score} incorporates imaging physics into score-based generation and uses range-null space decomposition to focus estimation on measurement-invisible components. 
DDMM-CT~\cite{li2025cross} further introduces cross-modal geometric priors to refine null-space components in sparse-view reconstruction while enforcing consistency with the acquired CT measurements.   
These approaches effectively confine generative modeling to poorly constrained image components, but their iterative reverse diffusion and repeated physics operations require substantial computational cost and long inference times. 

\subsubsection{Implicit Null-Space Estimation}

Another class of methods avoids explicit range-null space decomposition and instead constrains the sparse-view solution through missing view completion or iterative projection-image correction~\cite{jin2024staran,lin2024c,wang2022dudotrans,li2023mdst}.
Although missing projections are not equivalent to null-space components, complementary views can capture responses from structures that are invisible or weakly constrained by the acquired measurements. 
Selected complementary projections can therefore serve as low-dimensional coordinates that constrain image variations associated with the null space of the imaging operator.

According to the information interaction between the projection and image domains, these methods can be broadly grouped into three categories. 
(a) Projection completion with cascaded dual-domain refinement: these methods first complete the sparse-view sinogram, reconstruct an image through an analytical or learned operator, and then refine residual artifacts in the image domain, such as WNet~\cite{cheslerean2023wnet} and MVMS-RCN~\cite{fan2024mvms}. 
MVMS-RCN further combines multi-sparsity projection correction with multiscale image refinement to improve unified reconstruction across different numbers of views. 
(b) Residual-driven physics-consistency correction: these methods derive projection residuals from the mismatch between the current reconstruction and acquired measurements, and map them into image-domain corrections via backprojection, primal-dual updates, or learned error mappings. 
Representative approaches include DRONE~\cite{wu2021drone}, DREAM-Net~\cite{zhang2022dream}, and PAIL~\cite{zhang2025trustworthy}, Primal-Dual UNet~\cite{ernst2022primal}, and LIRE~\cite{moriakov2023end}. 
 (c) Cyclic interaction and alternating optimization of joint projection--image variables: these approaches jointly optimize sinogram and image through alternating projection restoration, image reconstruction, and physics-consistency updates. For example, LEARN++~\cite{zhang2022learn++} continuously exchanges features and intermediate estimates between the projection and image domains through a recursive dual-domain network. 
 LAMA~\cite{sun2025efficient} starts from an explicit dual-domain variational model and unfolds its alternating minimization procedure into a learnable network.

Despite these methods have achieved advanced perforemance for sparse-view reconstruction, they still have three major limitations. 
First, null-space estimation remains highly uncertain. Explicit approaches often target the full high-dimensional null-space component, which is weakly constrained by sparse measurements and therefore difficult to estimate reliably, potentially causing oversmoothing or unsupported structures. 
Second, heterogeneous errors are often entangled in implicit null-space estimation. Projection completion, residual correction, and dual-domain methods typically mix measurement errors with prediction errors despite their distinct origins, and their joint propagation through backprojection or iterative updates may introduce structured artifacts and progressively degrade reconstruction quality. 
Third, structural constraints on poorly observed components remain insufficient. Existing range-null space methods mainly enforce measurement consistency but provide limited structural regularization for null- and near-null-space components, allowing learned priors to introduce unsupported or anatomically inconsistent structures under severe undersampling or distribution shifts.

\subsection{Discretization Dependence}

Another challenge is the dependence of learning-based CT reconstruction on fixed discretizations. Models trained at specific angular sampling rates and image resolutions often generalize poorly to different acquisition or reconstruction grids. 
Recent studies have therefore explored unified reconstruction frameworks to accommodate multiple sampling configurations within a single model. 
For example, MVMS-RCN~\cite{fan2024mvms} alleviates sampling-rate dependence by using a unified dual-domain unfolding model to handle multiple view numbers. 
CvG-Diff~\cite{chen2025cross} models different angular sparsities as successive levels of a deterministic degradation process, enabling unified training across multiple sparsity levels. However, these methods primarily address variations in angular sampling rather than cross-resolution reconstruction. INR-based approaches provide continuous image representations and can naturally be evaluated at arbitrary spatial coordinates, but many require instance-specific optimization of the implicit network, resulting in substantial test-time computation~\cite{chen2023sparse, shi2024implicit, friis2025implicit, mohan2025distributed}.

Neural operators offer a more principled route to discretization-agnostic reconstruction by learning mappings between function spaces with parameters shared across discretizations~\cite{kovachki2023neural, ocampo2022scalable, liu2024neural, jatyani2025unified}. Recently,  CT neural Operator (CTO) employs U-shaped DISCO neural operators (UDNO) in both sinogram and image domains and exploits function-space parameterization to generalize across projection sampling rates and image resolutions without retraining~\cite{datta2025resolution}. It further incorporates angular periodicity and detector-domain frequency modeling to respect CT acquisition geometry.
Nevertheless, its sinogram-space DISCO blocks use a common operator configuration and a shared support radius for the sinogram domain. This design does not explicitly parameterize the distinct physical scales of angular and detector coordinates, motivating an anisotropic operator with independently controlled angular and detector supports.
\begin{figure*}[t] 
    \centering
    \includegraphics[width=0.95\textwidth]{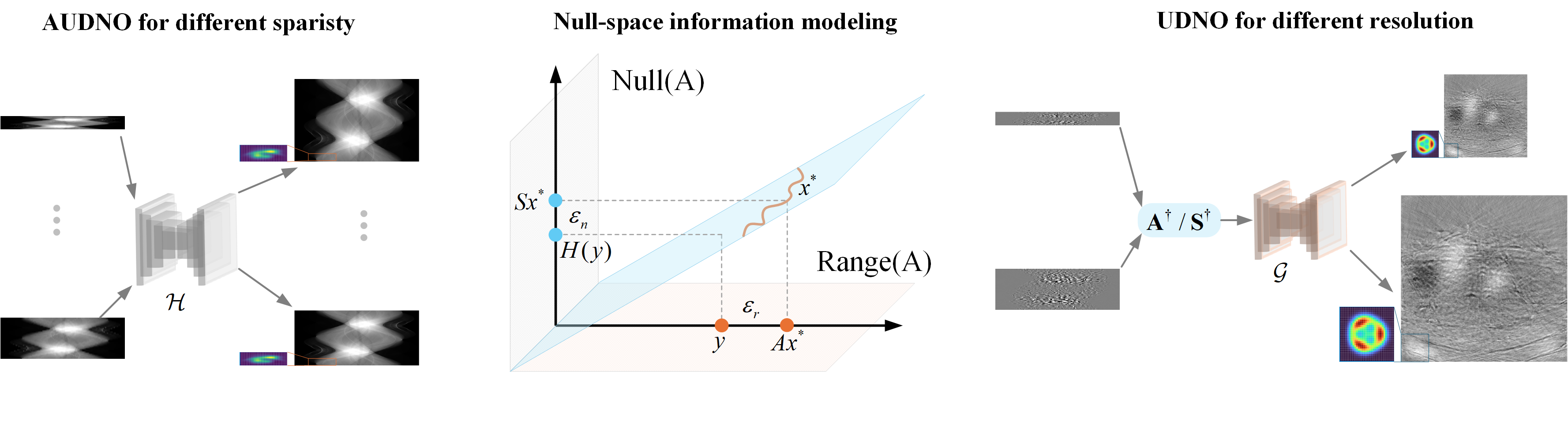} 
    \caption{Conceptual overview of SONAR. From left to right, the
sinogram-domain A-UDNO predicts null-space-aware pseudo-measurements
across different angular sampling densities; the dual-residual model
independently characterizes uncertainties in the acquired measurements and
learned pseudo-measurements; and the corresponding physics-lifted
residuals are processed by re-discretizable image-domain UDNOs to impose
structural consistency across reconstruction resolutions.}
    \label{fig:SONAR_arch_A}
\end{figure*}

\subsection{Our approach}

To address the above limitations, we propose SONAR, a
\emph{Structure-Consistent Neural Operator for Null-Space-Aware
Reconstruction}. As shown in Fig. \ref{fig:SONAR_arch_A}, SONAR integrates null-space-aware pseudo-measurement
learning, dual-residual structural consistency, and cross-discretization
neural operators within an optimization-inspired unrolled framework.
The main innovations are summarized as follows.

\textit{1) Null-Space-Aware Pseudo-Measurement Learning:}
Based on the low-dimensional null-space projection strategy of
NPN~\cite{jacome2026npn}, SONAR replaces direct recovery of the full
high-dimensional null-space component with prediction of a compact
subspace representation. An auxiliary representation operator captures
reconstruction-relevant image variations that are invisible or weakly
constrained by the acquired measurements, while a pretrained predictor
infers the corresponding representation from the measured projections.
The predicted representation supplies learnable pseudo-measurements for
poorly observable image components, thereby reducing the effective
complexity and uncertainty of null-space information estimation.

\textit{2) Dual-Residual Structural Consistency:}
Extending NPN, SONAR explicitly models uncertainty in the learned
pseudo-measurements. It introduces a measurement residual to capture
acquisition noise and forward-model mismatch and a pseudo-measurement
residual to account for prediction errors in the learned subspace
representation. Beyond enforcing numerical consistency in their
respective projection domains, SONAR lifts both residuals into the image
domain through the corresponding physics-based inverse operators and
constrains their structural effects using two independent neural
operators. This dual numerical--structural consistency allows the
residuals to absorb admissible errors while preventing them from
carrying anatomical information or producing structurally unsupported
artifacts.

\textit{3) Cross-Discretization Neural Operator Design:}
To decouple the learned mappings from fixed sampling grids, we develop
an anisotropic U-shaped neural operator (A-UDNO) for pseudo-measurement
prediction and two image-domain UDNOs for structural regularization.
A-UDNO assigns independent continuous supports to the periodic angular
dimension and the nonperiodic detector dimension, enabling the same
operator to process different angular sampling patterns. The
image-domain UDNOs parameterize their kernels on continuous spatial
coordinates and re-discretize them on different reconstruction grids.
Together, these operators enable a single model to generalize across
projection sparsities and image resolutions without modifying its core
parameters.

We evaluate SONAR on the simulated AAPM dataset and clinical MARS
photon-counting CT data across seen and unseen view settings, unseen
image resolutions, and different energy bins. The results demonstrate
consistent improvements in reconstruction accuracy, structural
reliability, and cross-discretization generalization over competing
methods.

\section{Related Theories}

\subsection{Range-Null Space Learning}

Let a linear imaging operator $\mathbf{A} \in \mathbb{R}^{M \times N}$ map an $N$-dimensional unknown signal $\mathbf{x} \in \mathbb{R}^{N}$ to an $M$-dimensional measurement space. The null space of $\mathbf{A}$ is defined as
\begin{equation}
\begin{aligned}
    \operatorname{Null}(\mathbf{A})
    &= \left\{ \mathbf{x} \in \mathbb{R}^N : \mathbf{A}\mathbf{x} = 0 \right\} \\
    &= \left\{ \mathbf{x} : \mathbf{x} \perp \mathbf{a}_m,\
    \ \forall m \in \{1,\ldots,M\} \right\}.
\end{aligned}
    \label{eq:null_space_definition}
\end{equation}
where $\mathbf{a}_m$ denotes the $m^{th}$ row vector of $\mathbf{A}$. Assuming that $\mathbf{A}$ has full row rank, let $\mathbf{A}^{\dagger} \in \mathbb{R}^{N \times M}$ denote its right pseudoinverse, such that $\mathbf{A}\mathbf{A}^{\dagger}=\mathbf{I}_M$. For any $\mathbf{x} \in \mathbb{R}^{N}$, there exists a unique decomposition
\begin{equation}
    \mathbf{x}
    = \mathbf{A}^{\dagger}\mathbf{A}\mathbf{x}
    + \left(\mathbf{I}-\mathbf{A}^{\dagger}\mathbf{A}\right)\mathbf{x},
    \label{eq:range_null_decomposition}
\end{equation}
where the first term belongs to the range-space component and the second term is the null-space component.

For sparse-view CT, the relationship between the unknown signal and the measurements can be expressed as
\begin{equation}
    \mathbf{y} = \mathbf{A}\mathbf{x} + \boldsymbol{\epsilon},
    \label{eq:forward_model_related}
\end{equation}
where $\boldsymbol{\epsilon}$ is the noise, and $M \ll N$. Substituting \eqref{eq:forward_model_related} into \eqref{eq:range_null_decomposition} yields
\begin{equation}
    \mathbf{x}
    = \mathbf{A}^{\dagger}\mathbf{y}
    - \mathbf{A}^{\dagger}\boldsymbol{\epsilon}
    + \left(\mathbf{I}-\mathbf{A}^{\dagger}\mathbf{A}\right)\mathbf{x}.
    \label{eq:range_null_decomposition_noisy}
\end{equation}
In Eq.\eqref{eq:range_null_decomposition_noisy}, both $\mathbf{x}$ and $\boldsymbol{\epsilon}$ are unknown. To solve this problem, DDN~\cite{chen2020deep} provides a solution by using one network to estimate the noise and another to estimate the null-space component:
\begin{equation}
\begin{split}
    \hat{\mathbf{x}}
    ={}& \mathbf{A}^{\dagger}\mathbf{y}
    - \mathbf{A}^{\dagger}\mathcal{R}_r\!\left(\mathbf{A}^{\dagger}\mathbf{y}\right) \\
    &+ \left(\mathbf{I}-\mathbf{A}^{\dagger}\mathbf{A}\right)
    \mathcal{R}_n\!\left(
        \mathbf{A}^{\dagger}\mathbf{y}
        - \mathbf{A}^{\dagger}\mathcal{R}_r\!\left(\mathbf{A}^{\dagger}\mathbf{y}\right)
    \right),
\end{split}
    \label{eq:ddn_reconstruction}
\end{equation}
where $\mathcal{R}_r$ fits the range-space noise residual and $\mathcal{R}_n$ estimates the complete null-space component. However, a key limitation of this type of methods is that the high-dimensional null-space component makes its direct estimation from limited measurements highly uncertain and prone to oversmoothing or unsupported structures. To address this issue, NPN~\cite{jacome2026npn} adopts a low-dimensional projection strategy:
\begin{equation}
    \hat{\mathbf{x}}
    = \mathbf{A}^{\dagger}\mathbf{y}
    + \gamma \mathbf{S}^{\dagger}\mathcal{G}(\mathbf{y}),
    \label{eq:npn_reconstruction}
\end{equation}
where $\mathbf{S}\in\mathbb{R}^{P\times N}$ is the projection matrix, which spans a structured low-dimensional subspace
orthogonal to the measurement operator $\mathbf{A}$, with
\begin{equation}
    \mathbf{s}_p \perp \mathbf{a}_m,
    \quad \forall p\in\{1,\ldots,P\},
    \quad \forall m\in\{1,\ldots,M\}.
    \label{eq:npn_subspace_condition}
\end{equation}
where $P<(N-M)$ denotes the number of selected key directions. The problem can also be written as
\begin{equation}
    \hat{\mathbf{x}}
    = \arg\min_{\mathbf{x}}
    \left\|\mathbf{A}\mathbf{x}-\mathbf{y}\right\|_{\epsilon}^{2}
    + \gamma\left\|\mathcal{G}^{*}(\mathbf{y})-\mathbf{S}\mathbf{x}\right\|_2^2,
    \label{eq:npn_optimization}
\end{equation}
where $\mathbf{S}\mathbf{x}$ represents the projection coefficients of $\mathbf{x}$ onto a low-dimensional subspace of the null space. Rather than learning the complete null-space image, NPN selects a small set of informative null-space directions. Its theoretical analysis further indicates that learning in a low-dimensional null-space subspace is more effective.

\subsection{Neural Operators}

Neural Operators (NOs) learn mappings between infinite-dimensional function spaces, which have demonstrated strong cross-resolution performance in tasks such as partial differential equations solving. DISCO is an effcient NO that defines the convolution kernel as a continuous function in function space and discretizes it only at the resolution of the input. Specifically, for a continuous kernel function $\kappa$ and an input function $g$ defined on a compact set $D\subset\mathbb{R}^d$, the continuous convolution is written as
\begin{equation}
    (\kappa * g)(v)
    = \int_D \kappa(v-u)g(u)\,du.
    \label{eq:continuous_convolution}
\end{equation}
Given discrete sampling points $\{u_j\}_{j=1}^J$ and the corresponding numerical quadrature weights $\{q_j\}_{j=1}^J$, the convolution at an output location $v_i$ can be approximated by
\begin{equation}
    (\kappa * g)(v_i)
    \approx \sum_{j=1}^J \kappa(v_i-u_j)g(u_j)q_j.
    \label{eq:discrete_convolution}
\end{equation}

The continuous kernel can further be parameterized as a linear combination of continuous basis functions,
\begin{equation}
    \kappa_{\theta}(\xi)
    = \sum_{\ell=1}^{L}\theta_{\ell}\psi_{\ell}(\xi),
    \label{eq:disco_kernel_parameterization}
\end{equation}
where $\{\psi_{\ell}\}_{\ell=1}^{L}$ are basis functions defined in the continuous function space and $\{\theta_{\ell}\}_{\ell=1}^{L}$ are learnable parameters. It follows that
\begin{equation}
\begin{split}
    (\kappa_{\theta} * g)(v_i)
    \approx \sum_{j=1}^{J}
    \left[
        \sum_{\ell=1}^{L}
        \theta_{\ell}\psi_{\ell}(v_i-u_j)
    \right]
    g(u_j)q_j.
\end{split}
    \label{eq:disco_discretization}
\end{equation}

Importantly, DISCO parameterizes the convolution kernel in continuous coordinates rather than as a fixed discrete stencil. Specifically, the kernel is expanded over continuous basis functions with learnable coefficients that are shared across discretizations. When the grid changes, the basis functions and quadrature weights are re-evaluated at the new sample locations, while the kernel retains a fixed support in the continuous domain. Consequently, DISCO discretizes the same local integral operator across resolutions. In contrast, a standard convolution with a fixed kernel size in grid points has a shrinking physical receptive field under grid refinement and, without resolution-dependent rescaling, approaches a pointwise mapping rather than a fixed-support integral operator.

CTO employs two UDNOs in the sinogram and image domains, coupled with physics-based data consistency to support reconstruction across different sampling rates and image resolutions. However, the detector and angular dimensions of a sinogram exhibit different physical characteristics: the former mainly reflects local spatial correlations, whereas the latter is periodic and captures inter-view dependencies. Applying isotropic DISCO kernels and shared multiscale processing to both dimensions may therefore be suboptimal, as it cannot explicitly model their different correlation scales and may further weaken already sparse angular information.

\begin{figure*}[t] 
    \centering
    \includegraphics[width=1.0\textwidth]{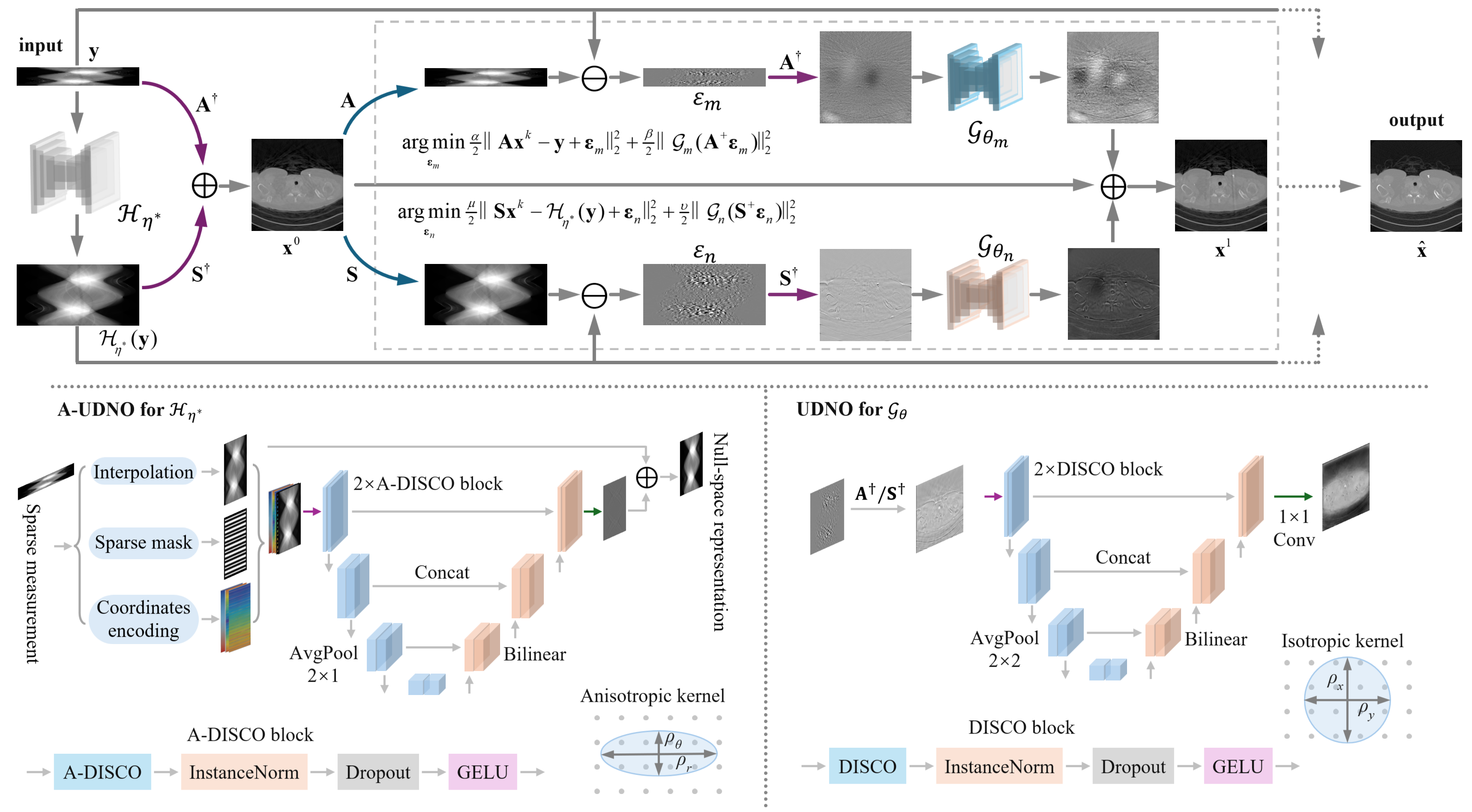} 
    \caption{Overall architecture of the proposed method. The pretrained
A-UDNO first completes the full-view projection while preserving the
measured views. At each cascade, the current reconstruction is forward
projected and compared with the measured and learned complementary
projections to form the range-space and null-space residuals, respectively.
The two residuals are backprojected and processed by independent
image-domain neural operators to update the reconstruction. The lower
panel shows (a) the A-UDNO with anisotropic DISCO convolutions in the sinogram
domain and (b) the image-domain UDNO based on standard DISCO convolutions.}
    \label{fig:SONAR_arch}
\end{figure*}

\section{Methodology}
\label{sec:method}


\subsection{Top-Level Design of SONAR}
\label{subsec:top_level}

To improve the reliability and discretization robustness of sparse-view
CT reconstruction, we propose SONAR (see Fig. \ref{fig:SONAR_arch}), which integrates two complementary
principles: null-space-aware dual-residual modeling and
cross-discretization neural operator learning.

First, inspired by the projected null-space modeling of
NPN~\cite{jacome2026npn}, SONAR avoids directly estimating the complete
high-dimensional null-space component in the image domain. Instead,
an auxiliary operator $\mathbf{S}$ represents reconstruction-relevant
information associated with image variations that are invisible or
poorly constrained by the measurement operator $\mathbf{A}$.
A pretrained predictor $\mathcal{H}_{\eta^\star}$ estimates the
reference representation $\mathbf{S}\mathbf{x}^{\star}$ from the
acquired measurements, thereby providing null-space-aware
pseudo-measurements. Predicting this structured representation reduces
the uncertainty associated with direct image-domain null-space
estimation.

Rather than treating the physical measurements and learned
pseudo-measurements as error-free constraints, SONAR explicitly models
their uncertainties. The measurement residual
$\boldsymbol{\epsilon}_m$ accounts for measurement noise and
forward-model mismatch, whereas the pseudo-measurement residual
$\boldsymbol{\epsilon}_n$ compensates for errors in
$\mathcal{H}_{\eta^\star}(\mathbf{y})$. In addition to enforcing
numerical agreement in their respective projection domains, the two
residuals are lifted into the image domain and regularized by two structural discriminator $\mathcal{G}_{\theta_m}$ and
$\mathcal{G}_{\theta_n}$,
independently. This dual-residual design allows admissible errors to be
absorbed while preventing them from explaining genuine anatomical
structures or introducing structurally unsupported content.

Second, all three learned mappings are implemented as neural operators
to reduce their dependence on fixed discretizations. The
sinogram-domain predictor $\mathcal{H}_{\eta^\star}$ is realized by an
anisotropic U-shaped neural operator (A-UDNO), which accounts for the
periodic angular geometry and the nonperiodic detector geometry and can
be re-discretized across different angular sampling patterns. The two
image-domain operators $\mathcal{G}_{\theta_m}$ and
$\mathcal{G}_{\theta_n}$ are implemented using UDNOs defined on
continuous spatial coordinates, allowing the same learned operators to
be evaluated on different reconstruction grids. These designs provide
cross-sparsity pseudo-measurement prediction and cross-resolution
structural regularization, respectively.

Combining these two principles, the proposed framework is formulated as
\begin{flalign}
&\operatorname*{arg\,min}_{\mathbf{x},\boldsymbol{\epsilon}_m,\boldsymbol{\epsilon}_n}\quad
\frac{\alpha}{2}
\left\|\mathbf{A}\mathbf{x}-\mathbf{y}+\boldsymbol{\epsilon}_m\right\|_2^2
+\frac{\beta}{2}
\left\|\mathcal{G}_{\theta_m}\!\left(\mathbf{A}^{\dagger}\boldsymbol{\epsilon}_m\right)\right\|_2^2
\nonumber\\
&+\frac{\mu}{2}
\left\|\mathbf{S}\mathbf{x}-\mathcal{H}_{\eta^*}(\mathbf{y})+\boldsymbol{\epsilon}_n\right\|_2^2
+\frac{\nu}{2}
\left\|\mathcal{G}_{\theta_n}\!\left(\mathbf{S}^{\dagger}\boldsymbol{\epsilon}_n\right)\right\|_2^2.
&& \label{eq:joint_objective}
\end{flalign}
Here, The parameters $\alpha$ and $\mu$ weight the two numerical-consistency terms, while $\beta$ and $\nu$ control the corresponding structural constraints. The generalized inverse operators $\mathbf{A}^{\dagger}$ and $\mathbf{S}^{\dagger}$ lift the two residuals into the image domain.

The first and third terms in Eq.~\eqref{eq:joint_objective} enforce numerical consistency with the acquired measurements and learned pseudo-measurements, respectively, whereas the second and fourth terms restrict the image-domain structures that can be explained by their residuals. 
By explicitly separating the two branches, the formulation prevents physical measurement uncertainty and learned prediction uncertainty from being conflated and allows each information source to contribute through an independent soft constraint. 

\subsection{Null-Space-Aware Dual-Residual Modeling }
\label{subsec:range_null_framework}

In the following, we present two core components of the
proposed approach. We first describe how to construct and learn
null-space-aware pseudo-measurements from the acquired data, and then
develop a dual-residual separation and structural-consistency mechanism
to disentangle measurement and prediction errors and constrain their
image-domain structural effects.

\subsubsection{Learning Null-Space-Aware Pseudo-Measurements }
\label{subsubsec:low_dim_null}

In Eq.~\eqref{eq:joint_objective}, $\mathbf{S}$ serves as an auxiliary representation operator to probe image variations associated with the null and near-null spaces of $\mathbf{A}$. For an unstructured sensing operator, orthonormal directions in $\operatorname{Null}(\mathbf{A})$ can be obtained through a full QR factorization of $\mathbf{A}^{\top}$ and used to construct $\mathbf{S}$. For sparse-view CT, we instead exploit the acquisition geometry. Let $\mathbf{R}$ denote the full-view projection operator, and let $\Omega$ and $\Omega^c$ denote the acquired and complementary angle sets, respectively. We define
\begin{equation}
\mathbf{A}=\mathbf{P}_{\Omega}\mathbf{R},
\qquad
\mathbf{S}=\mathbf{P}_{\Omega^c}\mathbf{R},
\label{eq:ct_operator_partition}
\end{equation}
where $\mathbf{P}_{\Omega}$ and $\mathbf{P}_{\Omega^c}$ are the corresponding angular selection operators. Consequently, $\mathbf{S}\mathbf{x}$ collects the complementary-view projections of $\mathbf{x}$. Rather than reconstructing the complete high-dimensional null-space component, the predictor $\mathcal{H}_{\eta}$ estimates the reference response $\mathbf{S}\mathbf{x}^*$ from the acquired measurements $\mathbf{y}$, thereby providing learned pseudo-measurements for the null-space-aware branch.

The CT-specific operator $\mathbf{S}$ should be interpreted as a physics-based, null-space-sensitive operator rather than an exact orthogonal projector onto $\operatorname{Null}(\mathbf{A})$. In particular, a perturbation $\mathbf{z}\in\operatorname{Null}(\mathbf{A})$ leaves the acquired measurements unchanged, i.e., $\mathbf{A}\mathbf{z}=0$, while it may produce a nonzero complementary response $\mathbf{S}\mathbf{z}\neq0$. The complementary views can therefore reveal part of the ambiguity invisible to $\mathbf{A}$ and provide additional constraints without explicitly constructing the full null-space projector. 

Given paired reference images $\mathbf{x}^*$ and measurements $\mathbf{y}$, $\mathcal{H}_{\eta}$ is pretrained by
\begin{equation}
\eta^*
=
\operatorname*{arg\,min}_{\eta}
\mathbb{E}_{(\mathbf{x}^*,\mathbf{y})}
\left[
\left\|
\mathcal{H}_{\eta}(\mathbf{y})
-
\mathbf{S}\mathbf{x}^*
\right\|_2^2
\right].
\label{eq:pseudo_measurement_pretraining}
\end{equation}
Eq.~\eqref{eq:pseudo_measurement_pretraining} specifies the target of the predictor but does not assume that $\mathcal{H}_{\eta^*}(\mathbf{y})$ is error-free. Although NPN~\cite{jacome2025npn} acknowledges prediction mismatch, its reconstruction objective directly penalizes the discrepancy between the predicted representation and $\mathbf{S}\mathbf{x}$. In contrast, our formulation explicitly models this discrepancy using $\epsilon_n$ and constrains its image-domain structural effect through
$\mathcal{G}_{\theta_n}(\mathbf{S}^{\dagger}\epsilon_n)$. This design allows admissible prediction errors to be absorbed while discouraging them from introducing unsupported image structures. Moreover, because the angular grids vary with the number and locations of acquired views, we implement $\mathcal{H}_{\eta}$ using A-UDNO, whose continuous kernels can be re-discretized on different sampling grids. This establishes the connection between null-space-aware pseudo-measurement learning and the cross-discretization neural operator design introduced subsequently.


\subsubsection{Dual-Residual Separation and Structural Consistency}
\label{subsubsec:error_separation}

The acquired measurements $\mathbf{y}$ and the learned pseudo-measurements $\mathcal{H}_{\eta^*}(\mathbf{y})$ have different uncertainty sources. Eq.~\eqref{eq:joint_objective} therefore introduces a measurement residual $\boldsymbol{\epsilon}_m$ and a pseudo-measurement residual $\boldsymbol{\epsilon}_n$, allowing the two information sources to impose independent soft constraints on the reconstruction.

Residual separation alone, however, is insufficient. For any fixed $\mathbf{x}$, choosing
\begin{equation}
\boldsymbol{\epsilon}_m
=
\mathbf{y}-\mathbf{A}\mathbf{x},
\qquad
\boldsymbol{\epsilon}_n
=
\mathcal{H}_{\eta^*}(\mathbf{y})-\mathbf{S}\mathbf{x}
\label{eq:unconstrained_residuals}
\end{equation}
eliminates both numerical discrepancies in Eq.~\eqref{eq:joint_objective}, allowing the residuals to absorb arbitrary inconsistencies, including genuine image structures. We therefore lift the residuals into the image domain:
\begin{equation}
\mathbf{d}_m
=
\mathbf{A}^{\dagger}\boldsymbol{\epsilon}_m,
\qquad
\mathbf{d}_n
=
\mathbf{S}^{\dagger}\boldsymbol{\epsilon}_n.
\label{eq:residual_lifting}
\end{equation}
Here, $\mathbf{d}_m$ and $\mathbf{d}_n$ represent the image-domain perturbations induced by measurement and pseudo-measurement inconsistencies, respectively. The independent operators $\mathcal{G}_{\theta_m}$ and $\mathcal{G}_{\theta_n}$ regularize their structural content according to the corresponding error statistics. Consequently, the residuals may accommodate admissible errors without freely explaining anatomical structures or introducing unsupported content.

The complementary roles of the two branches can be examined through the image-update subproblem. With the residuals fixed, the terms involving $\mathbf{x}$ reduce to
\begin{equation}
\min_{\mathbf{x}}\;
\frac{\alpha}{2}
\left\|
\mathbf{A}\mathbf{x}
-
(\mathbf{y}-\boldsymbol{\epsilon}_m)
\right\|_2^2
+
\frac{\mu}{2}
\left\|
\mathbf{S}\mathbf{x}
-
\bigl(
\mathcal{H}_{\eta^*}(\mathbf{y})
-
\boldsymbol{\epsilon}_n
\bigr)
\right\|_2^2 .
\label{eq:image_update_subproblem}
\end{equation}
The associated normal operator satisfies
\begin{equation}
\mathbf{M}
=
\alpha\mathbf{A}^{\top}\mathbf{A}
+
\mu\mathbf{S}^{\top}\mathbf{S},
\mathbf{v}^{\top}\mathbf{M}\mathbf{v}
=
\alpha\|\mathbf{A}\mathbf{v}\|_2^2
+
\mu\|\mathbf{S}\mathbf{v}\|_2^2 .
\label{eq:normal_operator}
\end{equation}
Thus, $\mathbf{S}$ provides additional curvature along directions that are weakly constrained by $\mathbf{A}$. On a target reconstruction subspace $\mathcal{X}$, the sufficient condition
\begin{equation}
\operatorname{Null}(\mathbf{A})
\cap
\operatorname{Null}(\mathbf{S})
\cap
\mathcal{X}
=
\{\mathbf{0}\}
\label{eq:joint_identifiability}
\end{equation}
ensures that no nonzero perturbation in $\mathcal{X}$ remains invisible to both branches. Although the complementary-view operator used in CT is not an exact null-space projector, this condition provides a useful interpretation: the pseudo-measurement branch reduces the joint ambiguity left by the acquired projections. Overall, $\mathbf{A}$ and $\mathbf{S}$ provide complementary constraints, $\boldsymbol{\epsilon}_m$ and $\boldsymbol{\epsilon}_n$ separate their uncertainties, and $\mathcal{G}_{\theta_m}$ and $\mathcal{G}_{\theta_n}$ restrict the structural content attributable to the two residuals.

\subsection{Cross-Discretization Neural Operator Design}
\label{subsec:operator_design}

The subsection III.B defines the roles of
$\mathcal{H}_{\eta}$,
$\mathcal{G}_{\theta_m}$, and
$\mathcal{G}_{\theta_n}$
in the dual-residual formulation. Here, we describe their domain-specific neural-operator implementations for different projection and image discretizations.

\subsubsection{Sinogram-Domain A-UDNO for $\mathcal{H}_{\eta}$}
\label{subsubsec:audno}

To accommodate variable angular sampling, we introduce an anisotropic U-shaped neural operator (A-UDNO) to implement $\mathcal{H}_{\eta}$. A sinogram is defined over the periodic angular domain and the nonperiodic detector domain, with each sampling location represented as
\begin{equation}
\mathbf{u}_i=(\theta_i,r_i)
\in
\mathbb{S}_{\theta}^{1}\times[-1,1]_r ,
\label{eq:projection_coordinate}
\end{equation}
where $\theta_i$ and $r_i$ denote the projection angle and normalized detector position, respectively. Angular periodicity is incorporated through the wrapped displacement
\begin{equation}
\Delta_{\theta}(\theta_i,\theta_j)
=
\operatorname{atan2}
\left[
\sin(\theta_j-\theta_i),
\cos(\theta_j-\theta_i)
\right].
\label{eq:periodic_displacement}
\end{equation}
Based on the continuous DISCO parameterization, A-DISCO defines the normalized local coordinate at scale $p$ as
\begin{equation}
\boldsymbol{\xi}_{ij}^{p}
=
\left(
\frac{\Delta_{\theta}(\theta_i,\theta_j)}{\rho_{\theta}},
\frac{r_j-r_i}{\rho_r}
\right),
\label{eq:anisotropic_coordinate}
\end{equation}
where $\rho_{\theta}$ and $\rho_r$ independently control the continuous supports along the angular and detector dimensions. The corresponding A-DISCO convolution is
\begin{equation}
(\mathcal{K}_p u)(\mathbf{u}_i)
\approx
\sum_{j\in\mathcal{N}_p(i)}
\sum_{\ell=1}^{L}
w_{\ell}^{p}
\psi_{\ell}(\boldsymbol{\xi}_{ij}^{p})
u(\mathbf{u}_j)q_j^{p},
\label{eq:adisco}
\end{equation}
where $\mathcal{N}_p(i)$ denotes the continuous-support neighborhood, $q_j^p$ is the quadrature weight, and $\psi_\ell$ and $w_\ell^p$ are the continuous basis functions and learnable coefficients, respectively. The wrapped angular displacement preserves continuity across angular boundaries, while the independent supports allow the kernel to capture the distinct correlation scales and coupled dependencies of the two sinogram dimensions.

To generate the learned pseudo-measurements, A-UDNO constructs three aligned inputs on the target angular grid: an interpolated sinogram $\mathbf{p}_0$, a binary acquisition mask $\mathbf{M}_{\Omega}$, and continuous angular--detector coordinate encodings. These inputs are concatenated and lifted into a latent feature space, followed by a U-shaped encoder--decoder in which each resolution level contains two A-DISCO blocks composed of A-DISCO, instance normalization, dropout, and GELU activation. The encoder applies $2\times1$ average pooling only along the angular dimension to aggregate cross-view information while preserving detector resolution. The decoder restores the angular resolution through bilinear interpolation and fuses multiscale encoder features through skip concatenations. A final $1\times1$ convolution predicts a residual correction, which is added to $\mathbf{p}_0$ through a global residual connection. The acquisition mask then restores the measured views, and the complementary component
$\mathcal{H}_{\eta}(\mathbf{y}_{\Omega})
=\mathbf{P}_{\Omega^c}\widehat{\mathbf{p}}$
is extracted as the learned pseudo-measurements. When the view configuration changes, only the sampling coordinates, neighborhoods, and quadrature weights are recomputed, allowing the same A-UDNO parameters to be reused across different angular discretizations.

\subsubsection{Image-Domain UDNOs for $\mathcal{G}_{\theta_m}$ and $\mathcal{G}_{\theta_n}$}
\label{subsubsec:image_udno}

The two image-domain structural operators share the same UDNO architecture but use independent parameters. Their inputs are the physics-lifted residuals
$\mathbf{d}_m=\mathbf{A}^{\dagger}\boldsymbol{\epsilon}_m$
and
$\mathbf{d}_n=\mathbf{S}^{\dagger}\boldsymbol{\epsilon}_n$,
and their outputs are
$\mathcal{G}_{\theta_m}(\mathbf{d}_m)$
and
$\mathcal{G}_{\theta_n}(\mathbf{d}_n)$,
respectively. Each UDNO adopts a U-shaped encoder--decoder in which every resolution level contains two standard DISCO blocks composed of DISCO, instance normalization, dropout, and GELU activation. Unlike A-UDNO, the encoder applies $2\times2$ average pooling along both spatial dimensions, while the decoder restores the spatial resolution through bilinear interpolation and combines encoder features using skip concatenations. A final $1\times1$ convolution maps the decoded features to the structural response. Because the two image axes share the same spatial geometry, standard DISCO kernels with isotropic continuous support are employed. When the image resolution changes, the same continuous kernels are evaluated on the new spatial grid by updating only the sampling coordinates and quadrature weights. Consequently, A-UDNO enables cross-sparsity pseudo-measurement prediction, whereas the two image-domain UDNOs provide cross-resolution structural regularization.

\subsection{Implementation Details}
The minimization problem in Eq.~\eqref{eq:joint_objective} admits
multiple solution strategies. One possible approach is to train the
three neural operators
$\mathcal{H}_{\eta}$,
$\mathcal{G}_{\theta_m}$, and
$\mathcal{G}_{\theta_n}$
separately and subsequently solve the resulting variational problem
using a conventional iterative optimizer. Howeer, such a decoupled procedure requires repeated inner iterations at inference and separates
operator learning from the reconstruction dynamics. We therefore
instantiate Eq.~\eqref{eq:joint_objective} as a $K$-stage
optimization-inspired unrolled network, enabling efficient finite-step
inference and end-to-end adaptation of two structural correction
operators.

\subsubsection{Optimization-Inspired Unrolling}
\label{subsubsec:optimization}

Because the pseudo-measurement predictor is pretrained and fixed,
$\mathbf{p}=\mathcal{H}_{\eta^*}(\mathbf{y})$ is computed once and
shared across all stages. Starting from an initial reconstruction
$\mathbf{x}^{0}$, the $k^{th}$ stage first evaluates the discrepancies in
the measurement and pseudo-measurement branches. Locally approximating
the nonlinear structural terms by quadratic surrogates yields the
following shrinkage-style residual updates:
\begin{equation}
\boldsymbol{\epsilon}_m^{k+1}
=
\frac{\mathbf{y}-\mathbf{A}\mathbf{x}^{k}}{1+\lambda_m},
\qquad
\boldsymbol{\epsilon}_n^{k+1}
=
\frac{\mathbf{p}-\mathbf{S}\mathbf{x}^{k}}{1+\lambda_n},
\label{eq:residual_update}
\end{equation}
where $\lambda_m=\beta/\alpha$ and
$\lambda_n=\nu/\mu$ control the residual allocation in the measurement
and pseudo-measurement branches, respectively. These parameters are learnable in the unrolled network.

Rather than exactly solving the coupled normal equation at every stage,
the residuals are lifted into the image domain as
$\mathbf{d}_m^{k+1}
=\mathbf{A}^{\dagger}\boldsymbol{\epsilon}_m^{k+1}$ and
$\mathbf{d}_n^{k+1}
=\mathbf{S}^{\dagger}\boldsymbol{\epsilon}_n^{k+1}$.
The corresponding image-domain UDNOs extract structure-consistent
components from these lifted residuals and transfer them to the current
reconstruction:
\begin{equation}
\mathbf{x}^{k+1}
=
\mathbf{x}^{k}
+\lambda_m
\mathcal{G}_{\theta_m}\!\left(\mathbf{d}_m^{k+1}\right)
+\lambda_n
\mathcal{G}_{\theta_n}\!\left(\mathbf{d}_n^{k+1}\right).
\label{eq:image_update}
\end{equation}
Thus, each stage successively performs residual estimation,
physics-based lifting, and structure-aware image correction. The two
UDNO branches use independent parameters to capture the distinct error
characteristics of the acquired measurements and learned
pseudo-measurements.

\subsubsection{Two-Stage Training}
\label{subsubsec:training}

Training proceeds in two stages. First, the sinogram-domain A-UDNO
$\mathcal{H}_{\eta}$ is pretrained to estimate the reference
representation $\mathbf{S}\mathbf{x}^{*}$ from the acquired
sparse-view measurements, as described in
Eq.~\eqref{eq:pseudo_measurement_pretraining}. Its predictions serve as
null-space-aware pseudo-measurements, and the optimized parameters
$\eta^*$ are subsequently frozen.

In the second stage, the $K$-stage dual-residual reconstruction network
is trained end-to-end while keeping
$\mathcal{H}_{\eta^*}$ fixed. Letting
$\widehat{\mathbf{x}}=\mathbf{x}^{K}$ denote the final reconstruction,
the training objective is
\begin{equation}
\mathcal{L}_{\mathrm{rec}}
=
\mathcal{L}_{1}
+\lambda_{\mathrm{str}}\mathcal{L}_{\mathrm{str}},
\label{eq:reconstruction_loss}
\end{equation}
where $\mathcal{L}_{1}$ enforces pixel-wise fidelity between
$\widehat{\mathbf{x}}$ and the reference image
$\mathbf{x}^{*}$, while $\mathcal{L}_{\mathrm{str}}$ combines
image-gradient and perceptual discrepancies. Freezing
$\mathcal{H}_{\eta^*}$ maintains a stable pseudo-measurement target,
allowing the remaining network parameters to focus on integrating the
two complementary information sources and suppressing structurally
implausible residual components.

\section{Experiment and Results}
\subsection{Experimental Setting}

\subsubsection{Dataset Preparation}

To evaluate the proposed method, extensive experiments are conducted on both a public CT dataset and a clinical PCCT dataset.

\textbf{AAPM Dataset:} The simulation study uses the AAPM Low-Dose CT Grand Challenge dataset, divided at the patient level: nine patients (4,795 slices) for training, one patient (318 slices) for validation, and one patient (823 slices) for testing. Full-view sinograms are synthesized from the reference images via a differentiable CT forward model matching the shared geometry. 
Sparse-view measurements are generated by uniformly subsampling the 373 views. Training is conducted on $6\times$, $8\times$, and $10\times$ downsampling factors (62, 47, and 37 views, treated as seen configurations). To evaluate zero-shot generalization across unseen sampling rates, models are evaluated at 41 and 31 views without fine-tuning.

\textbf{Clinical Wrist PCCT Dataset:}
The clinical study comprises wrist PCCT scans acquired on a MARS photon-counting system with a helical trajectory, rebinned into 2-D fan-beam geometry. Data from eight subjects are split at the subject level into six for training, one for validation, and one for testing. Scans are acquired at 120~kV, 35~$\mu$A, with 160~ms exposure per view across eight energy bins; the 1st (7--35~keV), 5th (55--60~keV), and 8th ($>70$~keV) bins are evaluated to benchmark performance across different noise and spectral conditions. 
Due to higher measurement noise and detector uncertainties in clinical PCCT, more moderate sparse-view configurations are adopted: 93 and 62 views ($\approx 4\times$ and $6\times$) are used as seen configurations during training, while 74 and 53 views serve as unseen configurations for zero-shot testing.

All framework is implemented in PyTorch and trained on an NVIDIA RTX A6000 GPU. To facilitate a controlled and consistent comparison between the simulation and clinical studies, both experiments are matched to the same CT acquisition geometry. Specifically, a source-to-isocenter distance (SOD) of 199.52~mm and a source-to-detector distance (SDD) of 271.88~mm are adopted. The detector had a physical pitch of 0.11~mm (1,547 elements for synthesized data). Full-view projections consisted of 373 uniformly distributed angles over a $360^{\circ}$ circular trajectory, and reference reconstructions are produced on a $256\times256$ grid (in-plane pixel size of 0.43~mm) using filtered backprojection (FBP). The reconstruction network is optimized using AdamW with an initial learning rate of $5\times10^{-4}$ and a weight decay of $1\times10^{-4}$. A cosine annealing schedule with a minimum learning rate of
$1\times10^{-6}$ is adopted. The batch size is 4, with gradient
accumulation used during training, and the network is trained for up to
100 epochs.

\begin{figure*}[htbp] 
    \centering
    \includegraphics[width=\textwidth]{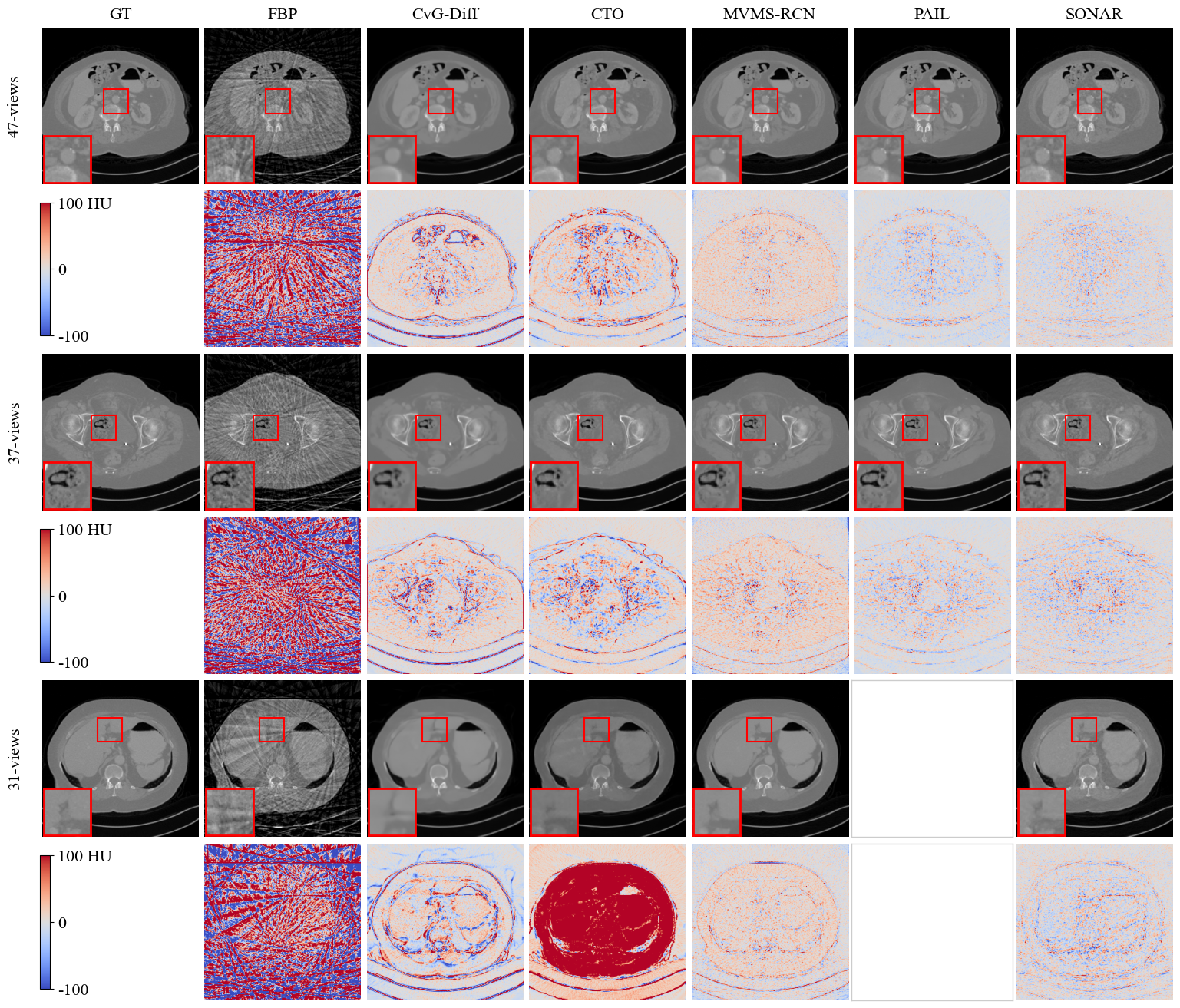} 
    \caption{Qualitative comparison of sparse-view CT reconstruction methods
on the AAPM dataset. Reconstructed images are displayed using a window of
$[-1000,1000]$ HU, and the corresponding error maps are displayed within
$[-100,100]$ HU. Red boxes indicate the regions of interest shown in the
magnified insets }
    \label{fig:seen_unseen_256}
\end{figure*}
\subsubsection{Compared Methods}

To guarantee a comprehensive evaluation, we benchmark \textbf{SONAR} against representative methods from complementary sparse-view CT reconstruction paradigms. \textbf{FBP} serves as the conventional analytical standard to quantify direct degradation from sparse angular sampling. \textbf{MVMS-RCN}~\cite{fan2024mvms} represents a unified dual-domain approach tailored for multi-sparsity configurations. \textbf{CvG-Diff}~\cite{chen2025cross} introduces a generative diffusion model explicitly developed for cross-view generalization. Serving as the most closely related neural-operator baseline, \textbf{CTO}~\cite{datta2025resolution} targets reconstruction across continuous projection rates and image resolutions, facilitating a direct assessment of our anisotropic operator design. Finally, \textbf{PAIL}~\cite{zhang2025trustworthy} exemplifies physics-guided unrolled networks that emphasize progressive artifact suppression, rigorously testing our dual-residual consistency mechanism. 

\subsubsection{Evalluation Metrics}

For quantitative evaluation, peak signal-to-noise ratio (PSNR),
structural similarity index measure (SSIM), and root mean square error
(RMSE) are adopted as the primary image-quality metrics. Higher PSNR and
SSIM values and a lower RMSE indicate better reconstruction performance.
All quantitative results are reported as the mean and standard deviation
over the corresponding test set.

To evaluate cross-sparsity generalization, we first report reconstruction
performance under the view configurations included during training,
referred to as the \emph{seen} settings. The trained models are then
directly evaluated under additional view configurations that are excluded
from training, referred to as the \emph{unseen} settings, without
parameter updating or task-specific retraining.

To evaluate cross-resolution generalization, the models trained at
$256\times256$ are directly applied to the target reconstruction grids of
$384\times384$ and $512\times512$. The reconstructed images are generated
directly on the target grids and quantitatively evaluated without
resolution-specific retraining, fine-tuning, or post-reconstruction
interpolation. This protocol therefore measures the zero-shot
generalization capability of each method across spatial discretizations.
For methods whose available implementations cannot be directly transferred
to an unseen view configuration or spatial resolution, the corresponding
results are denoted by ``--''.

\begin{figure}[htbp] 
    \centering
    \includegraphics[width=\columnwidth]{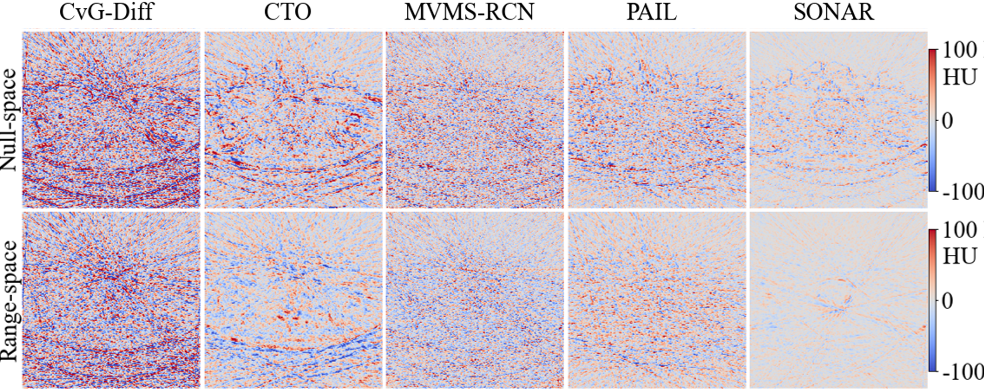} 
    \caption{error map of range-null space decomposition under the 47 views reconstruction on AAPM dataset. The display window is set to [-100,100] HU.}
    \label{fig:range_null_space_residual}
\end{figure}

\begin{figure}[t] 
    \centering
    \includegraphics[width=\columnwidth]{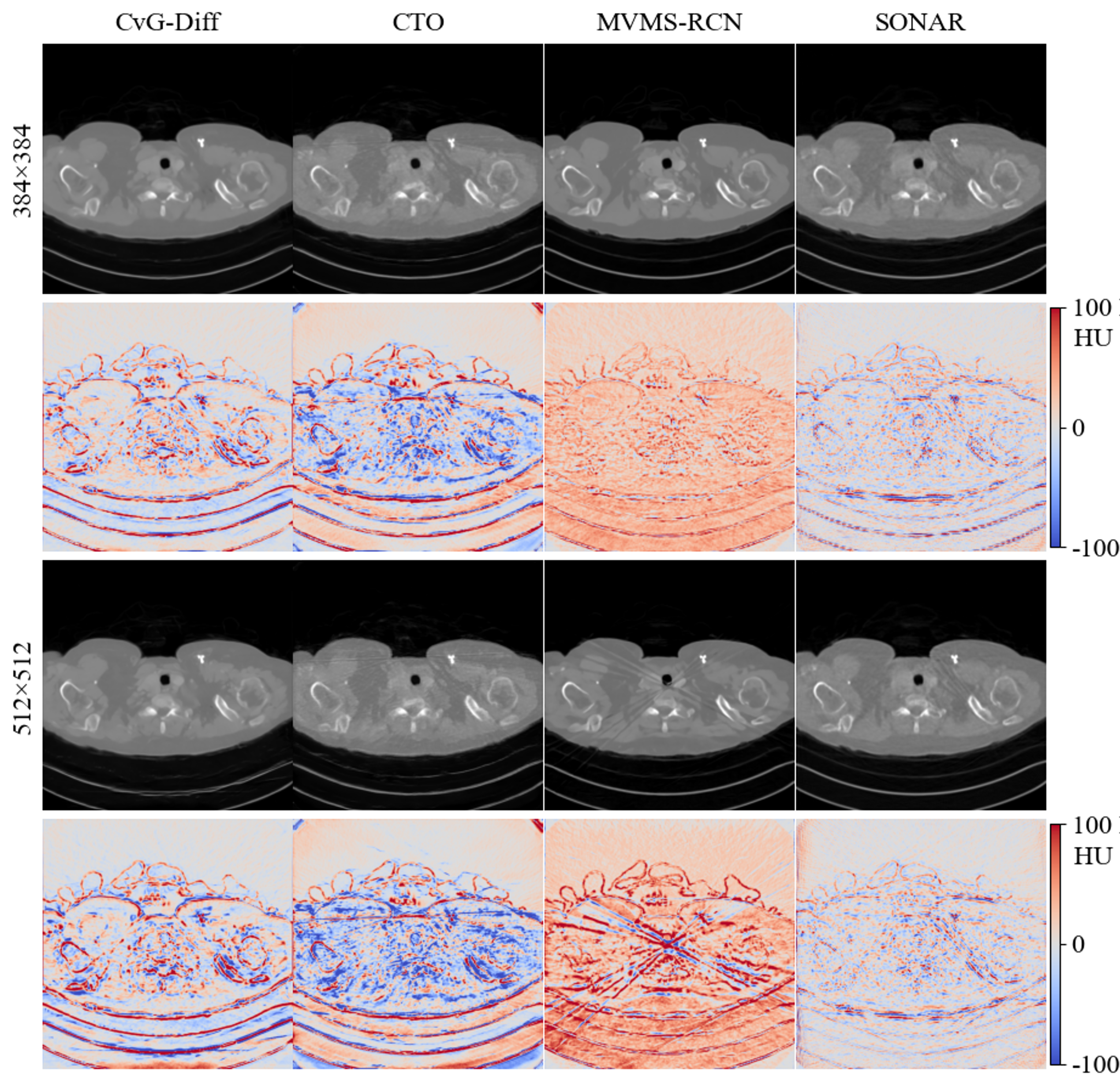} 
    \caption{Zero-shot cross-resolution reconstruction on the AAPM dataset.
All models are trained at $256\times256$ and directly evaluated at
$384\times384$ and $512\times512$ without resolution-specific
retraining. Error maps are displayed within $[-100,100]$ HU.}
    \label{fig:zero_shot_resolution}
\end{figure}

\begin{table*}[htbp]
\centering
\caption{Quantitative comparison under seen and unseen sparsity levels (Mean $\pm$ Std). Symbol $^*$ denotes sparsity levels not used during training and evaluated by direct zero-shot inference.}
\label{tab:seen_unseen_256_index}
\resizebox{\textwidth}{!}{
\begin{tabular}{c c c cccccc}
\toprule
\textbf{Setting} & \textbf{Views} & \textbf{Metric} & \textbf{FBP} & \textbf{CvG-Diff} & \textbf{CTO} & \textbf{MVMS-RCN} & \textbf{PAIL} & \textbf{SONAR} \\
\midrule
\multirow{9}{*}{\textbf{Seen}}
& \multirow{3}{*}{62}
 & PSNR$\uparrow$                      & $31.55 \pm 0.90$    & $41.55 \pm 1.26$    & $43.40 \pm 1.66$    & $45.48 \pm 1.01$   & $47.01 \pm 1.15$   & $\mathbf{48.88 \pm 1.64}$ \\
& 
 & SSIM$\uparrow$                      & $0.7085 \pm 0.0364$ & $0.9768 \pm 0.0064$ & $0.9770 \pm 0.0063$ & $0.9837 \pm 0.0036$& $0.9875 \pm 0.0030$ & $\mathbf{0.9908 \pm 0.0029}$ \\
& 
 & $\text{RMSE}_{\text{HU}}\downarrow$ & $108.89 \pm 11.24$  & $34.63 \pm 5.38$    & $28.21 \pm 5.73$    & $21.94 \pm 2.59$   & $18.43 \pm 2.54$   & $\mathbf{15.00 \pm 3.02}$ \\
\cmidrule{2-9}
& \multirow{3}{*}{47}
 & PSNR$\uparrow$                      & $26.81 \pm 0.72$    & $40.73 \pm 1.43$    & $42.01 \pm 1.73$    & $44.91 \pm 1.12$   & $46.69 \pm 1.76$   & $\mathbf{47.05 \pm 1.67}$ \\
& 
 & SSIM$\uparrow$                      & $0.5211 \pm 0.0485$ & $0.9706 \pm 0.0089$ & $0.9705 \pm 0.0081$ & $0.9813 \pm 0.0044$& $0.9856 \pm 0.0054$ & $\mathbf{0.9868 \pm 0.0041}$ \\
& 
 & $\text{RMSE}_{\text{HU}}\downarrow$ & $187.73 \pm 15.72$  & $38.18 \pm 6.68$    & $33.15 \pm 6.89$    & $23.48 \pm 3.10$   & $19.40 \pm 4.81$   & $\mathbf{18.53 \pm 3.71}$ \\
\cmidrule{2-9}
& \multirow{3}{*}{37}
 & PSNR$\uparrow$                      & $26.92 \pm 0.85$    & $40.41 \pm 1.52$    & $40.89 \pm 1.63$    & $44.31 \pm 1.26$   & $44.87 \pm 1.50$   & $\mathbf{45.41 \pm 1.71}$ \\
& 
 & SSIM$\uparrow$                      & $0.5116 \pm 0.0485$ & $0.9670 \pm 0.0101$ & $0.9642 \pm 0.0092$ & $0.9789 \pm 0.0056$& $0.9801 \pm 0.0055$ & $\mathbf{0.9818 \pm 0.0058}$ \\
& 
 & $\text{RMSE}_{\text{HU}}\downarrow$ & $185.60 \pm 18.38$  & $39.68 \pm 7.32$    & $37.64 \pm 7.28$    & $25.20 \pm 3.74$   & $23.73 \pm 4.26$   & $\mathbf{22.42 \pm 4.52}$ \\
\midrule
\multirow{6}{*}{\textbf{Unseen}}
& \multirow{3}{*}{$41^*$}
 & PSNR$\uparrow$                      & $26.76 \pm 0.79$    & $38.54 \pm 1.45$    & $41.14 \pm 1.58$    & $44.63 \pm 1.17$   & -                  & $\mathbf{45.23 \pm 1.39}$ \\
& 
 & SSIM$\uparrow$                      & $0.5115 \pm 0.0494$ & $0.9482 \pm 0.0117$ & $0.9660 \pm 0.0088$ & $\mathbf{0.9802 \pm 0.0049}$ & -        & $0.9800 \pm 0.0050$ \\
& 
 & $\text{RMSE}_{\text{HU}}\downarrow$ & $188.95 \pm 17.24$  & $49.17 \pm 8.47$    & $36.54 \pm 6.89$    & $24.26 \pm 3.32$   & -                  & $\mathbf{22.73 \pm 3.74}$ \\
\cmidrule{2-9}
& \multirow{3}{*}{$31^*$}
 & PSNR$\uparrow$                      & $27.02 \pm 0.88$    & $37.59 \pm 1.43$    & $31.04 \pm 0.76$    & $43.20 \pm 1.86$   & -                  & $\mathbf{43.21 \pm 1.39}$ \\
& 
 & SSIM$\uparrow$                      & $0.5172 \pm 0.0408$ & $0.9373 \pm 0.0141$ & $0.9310 \pm 0.0110$ & $\mathbf{0.9716 \pm 0.0135}$ & -        & $0.9703 \pm 0.0073$ \\
& 
 & $\text{RMSE}_{\text{HU}}\downarrow$ & $183.47 \pm 18.73$  & $54.80 \pm 9.37$    & $115.41 \pm 10.43$  & $29.01 \pm 6.62$   & -                  & $\mathbf{28.69 \pm 4.72}$ \\
\bottomrule
\end{tabular}
}
\end{table*}

\begin{table}[htbp]
\centering
\caption{Generalization and robustness evaluation across unseen image resolutions (Mean $\pm$ Std).}
\label{tab:zero_shot_resolution_index_AAPM}
\resizebox{\columnwidth}{!}{
\begin{tabular}{cc cccc}
\toprule
\textbf{Resolution} & \textbf{Metric} & \textbf{CvG-Diff} & \textbf{CTO} & \textbf{MVMS-RCN} & \textbf{SONAR} \\
\midrule
\multirow{3}{*}{$384 \times 384$} 
  & PSNR$\uparrow$                      & $41.08 \pm 1.22$    & $39.40 \pm 0.91$   & $41.50 \pm 1.19$    & $\mathbf{44.60 \pm 1.18}$ \\
  & SSIM$\uparrow$                      & $0.9575 \pm 0.0129$ & $0.9350 \pm 0.0114$& $0.9365 \pm 0.0189$ & $\mathbf{0.9774 \pm 0.0056}$ \\
  & $\text{RMSE}_{\text{HU}}\downarrow$ & $36.56 \pm 5.30$    & $44.15 \pm 4.62$   & $34.78 \pm 4.44$    & $\mathbf{24.34 \pm 3.40}$ \\
\midrule
\multirow{3}{*}{$512 \times 512$} 
  & PSNR$\uparrow$                      & $36.54 \pm 0.53$    & $37.63 \pm 0.77$   & $37.20 \pm 1.05$    & $\mathbf{45.26 \pm 1.45}$ \\
  & SSIM$\uparrow$                      & $0.9206 \pm 0.0167$ & $0.9049 \pm 0.0142$& $0.8718 \pm 0.0428$ & $\mathbf{0.9743 \pm 0.0073}$ \\
  & $\text{RMSE}_{\text{HU}}\downarrow$ & $61.09 \pm 3.78$    & $54.00 \pm 4.78$   & $56.97 \pm 6.77$    & $\mathbf{22.68 \pm 3.87}$ \\
\bottomrule
\end{tabular}
}
\end{table}

\subsection{Results on the AAPM Dataset}

\subsubsection{Performance on Seen View Settings}

Fig.~\ref{fig:seen_unseen_256} presents representative visual
comparisons at 47 and 37 views. Sparse-view FBP produces strong
directional streak artifacts and obscures low-contrast structures.
CvG-Diff and CTO suppress a large portion of these artifacts but retain
noticeable texture errors and local structural distortions. MVMS-RCN and
PAIL generate smoother reconstructions, although the smoothing attenuates
some fine anatomical details. In contrast, SONAR preserves local
boundaries and subtle structures more faithfully, while its error maps
contain weaker and less spatially correlated residual patterns.

Fig.~\ref{fig:range_null_space_residual} further decomposes the
47-view reconstruction error into range-space and null-space components.
The competing methods retain structured residuals in one or both
components. CvG-Diff exhibits pronounced high-frequency and streak-like
patterns, whereas CTO, MVMS-RCN, and PAIL reduce the overall error but
leave anatomically correlated structures, particularly in the null-space
component. SONAR suppresses structured errors in both components. The
reduced range-space error reflects improved agreement with the acquired
projections, while the weaker null-space error indicates more reliable
recovery of poorly observable image information.

Table~\ref{tab:seen_unseen_256_index} reports the quantitative results for
all three seen configurations. SONAR ranks first across PSNR, SSIM, and
RMSE. At 62 views, it improves PSNR by $1.87$~dB and reduces RMSE by
$3.43$~HU relative to PAIL, the strongest competing method under this
setting. The performance advantage persists at 47 and 37 views, confirming
stable reconstruction over the range of angular sampling densities used
during training.

\subsubsection{Performance on Unseen View Settings}
We next transfer the trained models directly to the unseen 41- and
31-view configurations without parameter updating. The released PAIL
implementation requires the testing sampling pattern to match its training
configuration; therefore, the comparison omits PAIL under these settings.

The 31-view example in Fig.~\ref{fig:seen_unseen_256} illustrates the
behavior under severe unseen undersampling. FBP contains extensive streak
artifacts, while several learning-based baselines retain structured errors
or distorted local details. CTO shows a particularly pronounced
reconstruction error under this setting. SONAR preserves the principal
anatomical structures and produces a more uniformly distributed,
lower-magnitude error map, despite receiving a sampling pattern absent
from training.

The quantitative results in
Table~\ref{tab:seen_unseen_256_index} support the visual comparison. At
41 views, SONAR improves PSNR by $0.60$~dB and reduces RMSE by
$1.53$~HU relative to MVMS-RCN, while both methods yield nearly identical
SSIM. At 31 views, SONAR and MVMS-RCN achieve comparable overall
performance: SONAR provides marginally higher PSNR and lower RMSE,
whereas MVMS-RCN obtains slightly higher SSIM. The absence of a marked
performance collapse under either unseen configuration indicates that
SONAR does not depend strongly on the discrete view numbers used during
training.

\subsubsection{Zero-Shot Cross-Resolution Generalization}

For cross-resolution evaluation, a model trained exclusively at
$256\times256$ directly generated reconstructions on
$384\times384$ and $512\times512$ grids. No method received
resolution-specific fine-tuning or post-reconstruction interpolation.
The released PAIL implementation requires retraining after changing the
spatial discretization and therefore does not appear in this comparison.

Fig.~\ref{fig:zero_shot_resolution} first compares the reconstructed
images and error maps on the two target grids. CvG-Diff retains noticeable
noise and boundary errors, while CTO leaves residual structural deviations.
MVMS-RCN shows substantial intensity and structural distortions,
particularly at $512\times512$. SONAR produces images that remain closer
to the reference, preserves anatomical boundaries more consistently, and
substantially reduces structured errors at both resolutions.

Table~\ref{tab:zero_shot_resolution_index_AAPM} confirms this advantage.
At $384\times384$, SONAR improves PSNR by $3.10$~dB and reduces RMSE
by $10.44$~HU relative to the strongest competing results. At
$512\times512$, the corresponding margins increase to $7.63$~dB and
$31.32$~HU, while SONAR also achieves the highest SSIM. These results
support the function-space formulation of SONAR: the learned continuous
operators can be re-discretized on a new image grid rather than transferred
as fixed-resolution convolutional weights.

\begin{figure*}[t] 
    \centering
    \includegraphics[width=\textwidth]{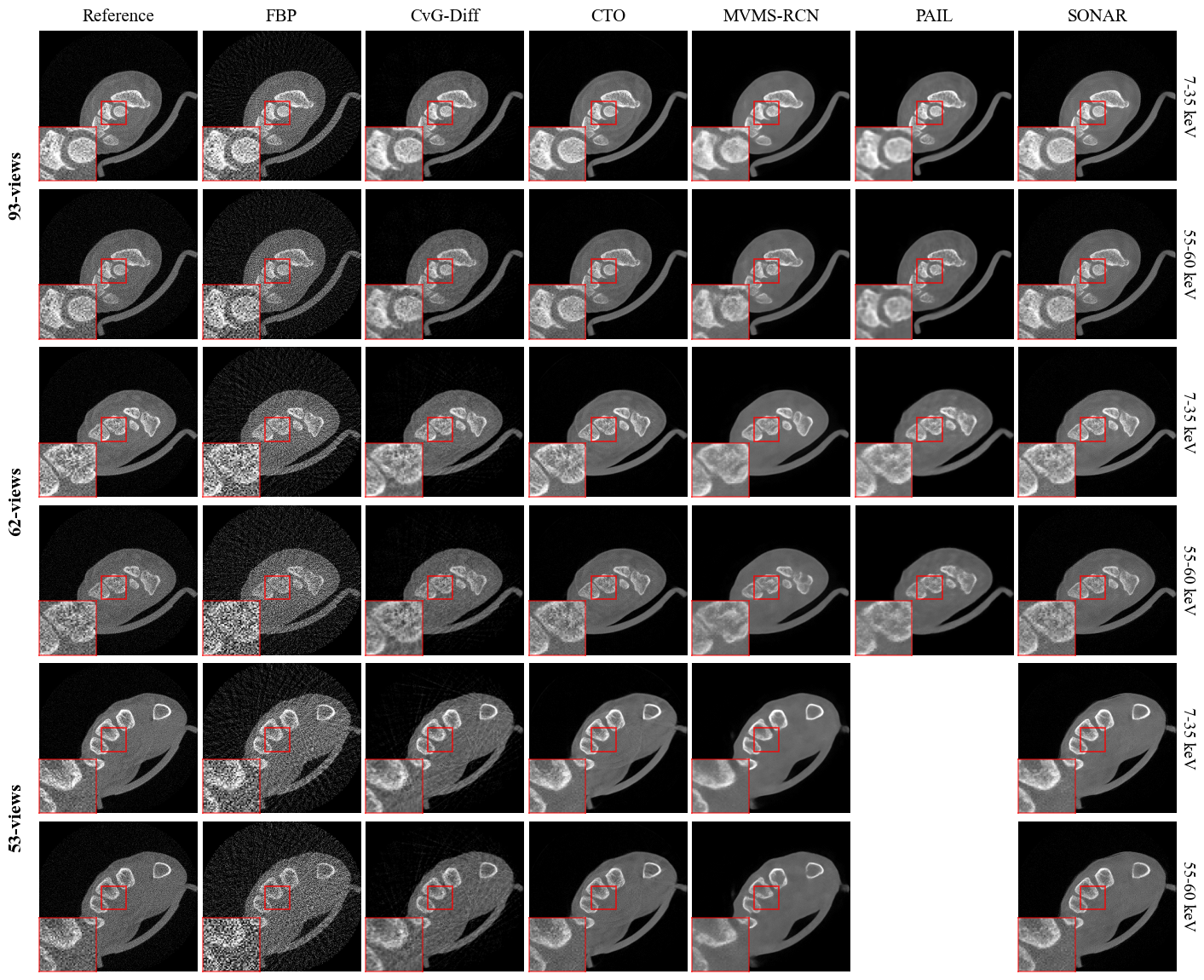} 
    \caption{Visual comparison of different reconstruction methods on the clinical wrist dataset under sparse-view settings. The green box indicates the region of interest (ROI). Results are evaluated under both seen (93,62 views) and unseen (53 views) sampling conditions during training.}
    \label{fig:clinical_virtual_result}
\end{figure*}

\begin{table*}[htbp]
\centering
\caption{Quantitative comparison under different sparsity levels and energy bins (Mean $\pm$ Std).}
\label{tab:clinical_Quantitative_result}
\resizebox{\textwidth}{!}{
\begin{tabular}{cc cccccc c}
\toprule
\textbf{Views} & \textbf{Metric} & \textbf{FBP} & \textbf{CvG-Diff} & \textbf{CTO} & \textbf{MVMS-RCN} & \textbf{PAIL} & \textbf{SONAR} & \textbf{Energy Bin} \\
\midrule
\multirow{6}{*}{93} 
 & PSNR$\uparrow$   & $26.61 \pm 0.25$    & $32.68 \pm 0.29$    & $32.34 \pm 0.31$    & $32.66 \pm 0.37$    & $29.22 \pm 0.22$    & $\mathbf{34.11 \pm 0.35}$ & \multirow{3}{*}{7-35keV} \\
 & SSIM$\uparrow$   & $0.480 \pm 0.005$   & $0.713 \pm 0.009$   & $0.774 \pm 0.012$   & $0.648 \pm 0.008$   & $0.549 \pm 0.006$   & $\mathbf{0.783 \pm 0.009}$ & \\
 & RMSE$\downarrow$ & $0.0467 \pm 0.0014$ & $0.0233 \pm 0.0008$ & $0.0242 \pm 0.0009$ & $0.0233 \pm 0.0010$ & $0.0346 \pm 0.0009$ & $\mathbf{0.0197 \pm 0.0008}$ & \\
\cmidrule{2-9}
 & PSNR$\uparrow$   & $24.22 \pm 0.20$    & $30.27 \pm 0.22$    & $30.51 \pm 0.23$    & $30.16 \pm 0.25$    & $28.20 \pm 0.16$    & $\mathbf{31.54 \pm 0.23}$ & \multirow{3}{*}{55-60keV} \\
 & SSIM$\uparrow$   & $0.404 \pm 0.004$   & $0.601 \pm 0.008$   & $0.694 \pm 0.006$   & $0.506 \pm 0.005$   & $0.425 \pm 0.003$   & $\mathbf{0.727 \pm 0.007}$ & \\
 & RMSE$\downarrow$ & $0.0615 \pm 0.0014$ & $0.0307 \pm 0.0008$ & $0.0298 \pm 0.0008$ & $0.0311 \pm 0.0009$ & $0.0389 \pm 0.0007$ & $\mathbf{0.0265 \pm 0.0007}$ & \\
\midrule
\multirow{6}{*}{74*} 
 & PSNR$\uparrow$   & $24.41 \pm 0.71$    & $28.88 \pm 0.05$    & $31.55 \pm 1.76$    & $31.82 \pm 0.85$    & -                    & $\mathbf{32.33 \pm 0.79}$ & \multirow{3}{*}{7-35keV} \\
 & SSIM$\uparrow$   & $0.395 \pm 0.014$   & $0.640 \pm 0.003$   & $0.654 \pm 0.016$   & $0.634 \pm 0.013$   & -                    & $\mathbf{0.725 \pm 0.088}$ & \\
 & RMSE$\downarrow$ & $0.0604 \pm 0.0049$ & $0.0360 \pm 0.0002$ & $0.0270 \pm 0.0056$ & $0.0258 \pm 0.0025$ & -                    & $\mathbf{0.0243 \pm 0.0022}$ & \\
\cmidrule{2-9}
 & PSNR$\uparrow$   & $22.34 \pm 0.63$    & $28.06 \pm 0.05$    & $29.80 \pm 1.44$    & $29.48 \pm 0.71$    & -                    & $\mathbf{29.93 \pm 0.67}$ & \multirow{3}{*}{55-60keV} \\
 & SSIM$\uparrow$   & $0.330 \pm 0.010$   & $0.539 \pm 0.003$   & $0.512 \pm 0.010$   & $0.497 \pm 0.009$   & -                    & $\mathbf{0.685 \pm 0.085}$ & \\
 & RMSE$\downarrow$ & $0.0766 \pm 0.0056$ & $0.0395 \pm 0.0002$ & $0.0328 \pm 0.0054$ & $0.0337 \pm 0.0027$ & -                    & $\mathbf{0.0320 \pm 0.0025}$ & \\
\midrule
\multirow{6}{*}{62} 
 & PSNR$\uparrow$   & $24.83 \pm 0.37$    & $32.14 \pm 0.27$    & $31.89 \pm 0.36$    & $32.52 \pm 0.40$    & $29.08 \pm 0.26$    & $\mathbf{33.50 \pm 0.36}$ & \multirow{3}{*}{7-35keV} \\
 & SSIM$\uparrow$   & $0.406 \pm 0.011$   & $0.713 \pm 0.007$   & $0.741 \pm 0.010$   & $0.641 \pm 0.008$   & $0.549 \pm 0.008$   & $\mathbf{0.745 \pm 0.015}$ & \\
 & RMSE$\downarrow$ & $0.0574 \pm 0.0025$ & $0.0247 \pm 0.0008$ & $0.0255 \pm 0.0011$ & $0.0237 \pm 0.0011$ & $0.0352 \pm 0.0010$ & $\mathbf{0.0212 \pm 0.0009}$ & \\
\cmidrule{2-9}
 & PSNR$\uparrow$   & $22.61 \pm 0.34$    & $29.96 \pm 0.21$    & $30.11 \pm 0.26$    & $30.08 \pm 0.26$    & $28.17 \pm 0.20$    & $\mathbf{31.04 \pm 0.25}$ & \multirow{3}{*}{55-60keV} \\
 & SSIM$\uparrow$   & $0.341 \pm 0.009$   & $0.604 \pm 0.007$   & $0.650 \pm 0.010$   & $0.499 \pm 0.005$   & $0.429 \pm 0.005$   & $\mathbf{0.690 \pm 0.011}$ & \\
 & RMSE$\downarrow$ & $0.0741 \pm 0.0029$ & $0.0318 \pm 0.0008$ & $0.0312 \pm 0.0010$ & $0.0314 \pm 0.0010$ & $0.0390 \pm 0.0009$ & $\mathbf{0.0281 \pm 0.0008}$ & \\
\midrule
\multirow{6}{*}{53*} 
 & PSNR$\uparrow$   & $23.27 \pm 0.68$    & $30.94 \pm 0.68$    & $31.45 \pm 1.64$    & $31.67 \pm 0.89$    & -                    & $\mathbf{33.15 \pm 0.74}$ & \multirow{3}{*}{7-35keV} \\
 & SSIM$\uparrow$   & $0.357 \pm 0.011$   & $0.687 \pm 0.019$   & $0.725 \pm 0.084$   & $0.628 \pm 0.014$   & -                    & $\mathbf{0.747 \pm 0.018}$ & \\
 & RMSE$\downarrow$ & $0.0688 \pm 0.0054$ & $0.0285 \pm 0.0022$ & $0.0272 \pm 0.0052$ & $0.0262 \pm 0.0027$ & -                    & $\mathbf{0.0221 \pm 0.0019}$ & \\
\cmidrule{2-9}
 & PSNR$\uparrow$   & $21.23 \pm 0.60$    & $29.06 \pm 0.63$    & $29.82 \pm 1.27$    & $29.38 \pm 0.74$    & -                    & $\mathbf{30.70 \pm 0.66}$ & \multirow{3}{*}{55-60keV} \\
 & SSIM$\uparrow$   & $0.301 \pm 0.007$   & $0.586 \pm 0.018$   & $0.656 \pm 0.015$   & $0.491 \pm 0.010$   & -                    & $\mathbf{0.691 \pm 0.079}$ & \\
 & RMSE$\downarrow$ & $0.0870 \pm 0.0061$ & $0.0353 \pm 0.0026$ & $0.0326 \pm 0.0048$ & $0.0341 \pm 0.0029$ & -                    & $\mathbf{0.0292 \pm 0.0022}$ & \\
\bottomrule
\end{tabular}
}
\end{table*}

\begin{figure}[htbp] 
    \centering
    \includegraphics[width=\columnwidth]{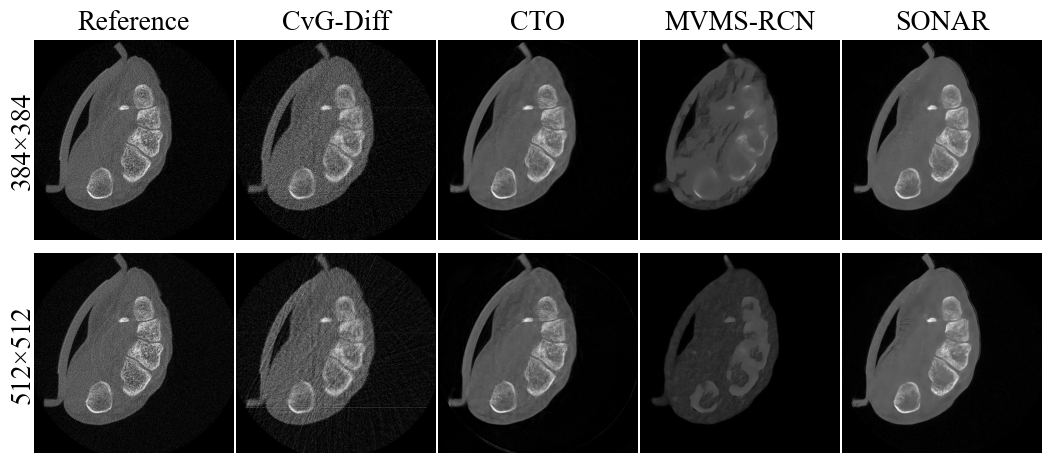} 
    \caption{Zero-shot cross-resolution reconstruction on the AAPM dataset.
All models are trained at $256\times256$ and directly evaluated at
$384\times384$ and $512\times512$ without resolution-specific
retraining.}
    \label{fig:zero-shot-super-resolution-pcct}
\end{figure}

\begin{table}[htbp]
\centering
\caption{Generalization and robustness evaluation across unseen image resolutions (Mean $\pm$ Std).}
\label{tab:zero_shot_resolution_index}
\resizebox{\columnwidth}{!}{
\setlength{\tabcolsep}{3pt}
\begin{tabular}{llcccc}
\toprule
\textbf{Resolution} & \textbf{Metric} & \textbf{CvG-Diff} & \textbf{CTO} & \textbf{MVMS-RCN} & \textbf{SONAR} \\
\midrule
\multirow{3}{*}{$384 \times 384$} 
  & PSNR$\uparrow$          & $25.24 \pm 1.53$    & $28.58 \pm 1.62$    & $24.98 \pm 0.95$    & $\mathbf{28.90 \pm 1.54}$ \\
  & SSIM$\uparrow$          & $0.507 \pm 0.040$   & $0.485 \pm 0.066$   & $0.345 \pm 0.056$   & $\mathbf{0.561 \pm 0.053}$ \\
  & $\text{RMSE}\downarrow$ & $0.0556 \pm 0.0101$ & $0.0379 \pm 0.0073$ & $0.0567 \pm 0.0063$ & $\mathbf{0.0365 \pm 0.0067}$ \\
\midrule
\multirow{3}{*}{$512 \times 512$} 
  & PSNR$\uparrow$          & $22.62 \pm 1.56$    & $25.90 \pm 1.64$    & $21.71 \pm 0.83$    & $\mathbf{26.31 \pm 1.58}$ \\
  & SSIM$\uparrow$          & $0.414 \pm 0.026$   & $0.377 \pm 0.053$   & $0.233 \pm 0.034$   & $\mathbf{0.457 \pm 0.042}$ \\
  & $\text{RMSE}\downarrow$ & $0.0752 \pm 0.0143$ & $0.0516 \pm 0.0101$ & $0.0825 \pm 0.0080$ & $\mathbf{0.0492 \pm 0.0092}$ \\
\bottomrule
\end{tabular}
}
\end{table}

\subsection{Results on Clinical PCCT Dataset}
\subsubsection{Performance on Seen View Settings}

Fig.~\ref{fig:clinical_virtual_result} compares representative
reconstructions from the 7--35~keV and 55--60~keV energy bins under the
seen sampling conditions. Sparse-view FBP retains substantial quantum
noise and angular undersampling artifacts, which obscure cortical
boundaries and internal bone structures. CvG-Diff and CTO suppress part
of the degradation but leave residual noise and nonuniform textures.
PAIL and MVMS-RCN produce smoother images, although some fine osseous
details become less distinct. SONAR provides a more favorable balance
between noise suppression and detail preservation, yielding clearer
cortical contours and more faithful internal structures in the magnified
regions.

Table~\ref{tab:clinical_Quantitative_result} reports consistent
improvements across both energy bins. At 93 views, SONAR improves PSNR
over the strongest competing methods by $1.43$~dB for the 7--35~keV
bin and by $1.03$~dB for the 55--60~keV bin. At 62 views, the
corresponding gains remain $0.98$ and $0.93$~dB, respectively, together
with the lowest RMSE values. Although the 55--60~keV measurements produce
lower overall image quality for all methods, SONAR retains its relative
advantage, demonstrating robustness to energy-dependent noise and spectral
variations.

\subsubsection{Performance on Unseen View Settings}

The trained models are further evaluated at 74 and 53 views, neither of
which occurs during training. The released PAIL implementation requires
a sampling configuration matched to training and therefore does not
provide results for these two settings.

The unseen 53-view examples in
Fig.~\ref{fig:clinical_virtual_result} show that severe undersampling and
real PCCT noise jointly challenge the reconstruction. FBP produces strong
noise and poorly defined bone structures, while the learning-based
baselines retain residual noise, excessive smoothing, or local structural
bias. SONAR better preserves cortical boundaries and internal bone
patterns and maintains a visual appearance closer to the full-view
reference in both energy bins.

Table~\ref{tab:clinical_Quantitative_result} shows that SONAR leads the
applicable methods at both unseen sampling densities. Under the more
sparsely sampled 53-view condition, SONAR improves PSNR by $1.48$~dB
for the 7--35~keV bin and by $0.88$~dB for the 55--60~keV bin relative
to the strongest baselines, while also producing the lowest RMSE.
The consistent performance at 74 and 53 views indicates that the learned
operator transfers to clinical sampling densities not represented during
training.

\subsubsection{Zero-Shot Cross-Resolution Generalization}

A model trained at $256\times256$ directly reconstructed clinical PCCT
images on $384\times384$ and $512\times512$ grids without
resolution-specific retraining or fine-tuning. The comparison excludes
PAIL because its released implementation requires a fixed training and
testing discretization.

Fig.~\ref{fig:zero-shot-super-resolution-pcct} compares the resulting
images with the corresponding full-view references. CvG-Diff retains
visible noise and streak-like structures, while CTO suppresses noise at
the cost of some local detail. MVMS-RCN exhibits stronger intensity and
structural distortions after transfer to the finer grids, especially at
$512\times512$. SONAR preserves the global anatomy, cortical boundaries,
and internal bone structures more faithfully at both target resolutions.

As reported in Table~\ref{tab:zero_shot_resolution_index}, SONAR achieves
the best overall quantitative performance on both grids. Relative to the
strongest competing method, it improves PSNR by $0.32$~dB at
$384\times384$ and by $0.41$~dB at $512\times512$, while also achieving
the highest SSIM and lowest RMSE. These results demonstrate that SONAR
can re-discretize its learned function-space representation on finer
clinical reconstruction grids without modifying the network parameters.
\subsection{Ablation Study}

\subsubsection{Effect of the Number of Cascades}

\begin{figure}[htbp] 
    \centering
    \includegraphics[width=\columnwidth]{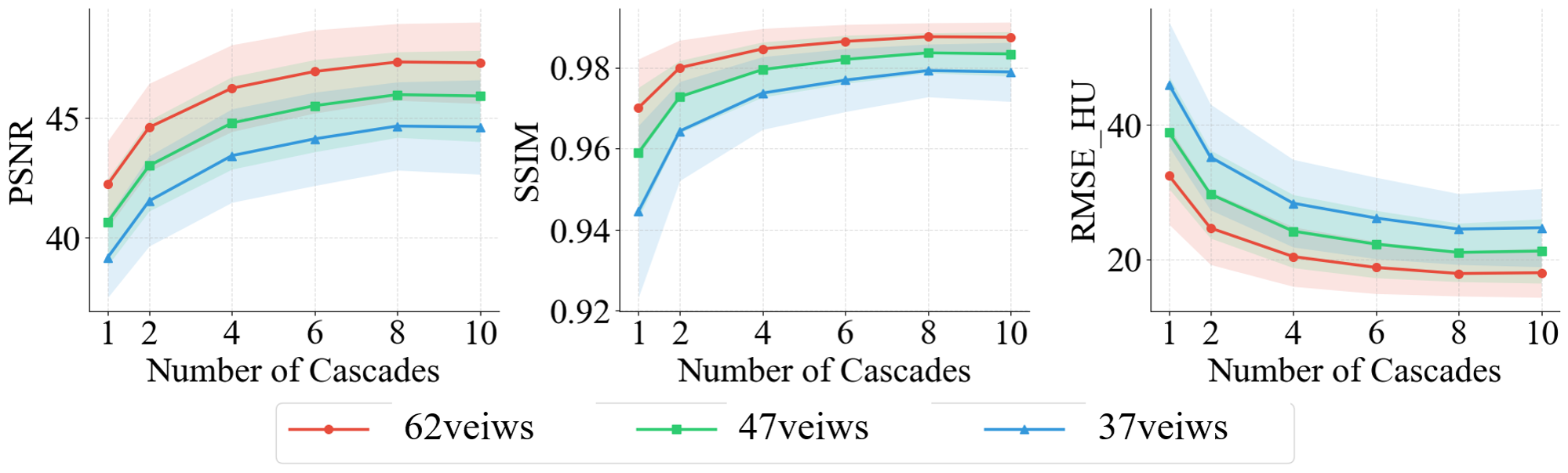} 
    \caption{Effect of the number of cascades on reconstruction performance
under the 62-, 47-, and 37-view settings. Performance improves
progressively as the number of cascades increases and becomes nearly
saturated at eight cascades.}
    \label{fig:cascade_ablation}
\end{figure}

We investigated the effect of the number of cascades under the 62-, 47-,
and 37-view settings. As shown in Fig.~\ref{fig:cascade_ablation},
increasing the number of cascades consistently improves reconstruction
performance across all three sampling densities. PSNR and SSIM gradually
increase, whereas RMSE decreases as the number of cascades increases
from 1 to 8, indicating that successive cascades progressively refine the
reconstruction.

The improvement becomes marginal when the number of cascades is further
increased from 8 to 10. The results obtained with 8 and 10 cascades are
nearly identical under all three view settings, suggesting that the
reconstruction performance has largely saturated at 8 cascades. Because
two additional cascades increase the computational and training costs
without providing a meaningful performance gain, eight cascades are used
in all subsequent experiments as a favorable trade-off between
reconstruction accuracy and computational efficiency.

\subsubsection{Effect of A-UDNO in sinogram domain}
To isolate the contribution of the proposed anisotropic design, both
A-UDNO and the conventional isotropic UDNO are trained exclusively
using the 47-view configuration while all other settings are kept
unchanged. The two variants are then evaluated at 47 views and directly
transferred to 37 and 62 views. The latter two configurations are unseen
with respect to this controlled single-view training protocol, although
they are included in the multi-sparsity training of the complete SONAR
model. The corresponding pretrained operators are also incorporated
into the full reconstruction framework to examine how their cross-view
generalization affects the final reconstruction performance.

\begin{figure}[htbp] 
    \centering
    \includegraphics[width=\columnwidth]{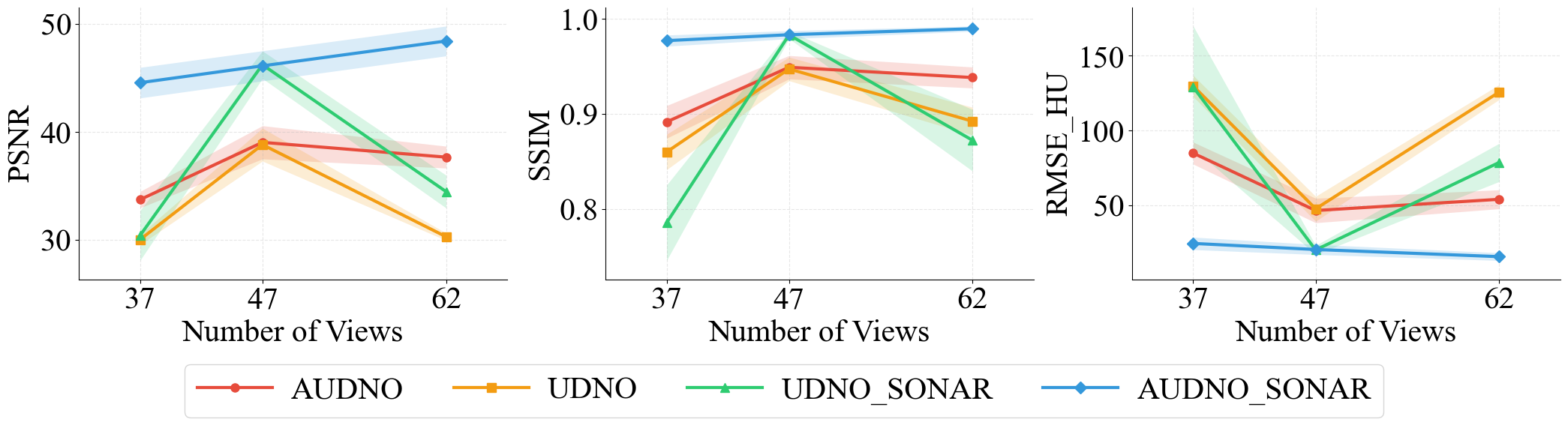} 
    \caption{Ablation study of the anisotropic neural operator under different view settings.}
    \label{fig:neural_operator_ablation}
\end{figure}

As shown in Fig.~\ref{fig:neural_operator_ablation}, A-UDNO and UDNO achieve
comparable performance at the view setting used for training, indicating
that both operators are capable of fitting the projection completion task
under a fixed sampling configuration. However, a pronounced difference
emerges when the sampling density deviates from the training condition.
The performance of UDNO decreases substantially at both 37 and 62 views,
as reflected by the marked reduction in PSNR and SSIM and the increase in
RMSE. In contrast, A-UDNO maintains considerably more stable performance
across different numbers of views.

This difference becomes even more evident after the pretrained operators
are incorporated into the complete reconstruction framework. The model
using A-UDNO maintains consistently high reconstruction accuracy when
transferred to unseen view settings, whereas replacing A-UDNO with UDNO
results in substantial performance degradation away from the training
sampling density. These results indicate that the performance gain does
not simply arise from increased fitting capacity at the training
configuration. Instead, explicitly accounting for the distinct angular
and detector-domain geometry enables A-UDNO to learn a more
sampling-robust continuous representation, which is critical for
generalization across different sparsity levels. This ablation therefore
validates the effectiveness of the proposed anisotropic continuous
operator design.

\subsection{Computational Complexity}
\begin{table}[htbp]
  \centering
  \caption{THE PARAMETER SIZE AND TEST TIME OF DIFFERENT METHODS WITH AN NVIDIA A6000 GPU}
  \label{tab:complexity of model}
    \begin{tabular}{lcc}
    \toprule
    Method & Testing time(s) & Parameters(M) \\
    \midrule
    CTO & 0.40 & 172 \\
    CvG\_Diff & 0.89 & 1.85 \\
    MVMS-RCN & 0.05 & 0.29 \\
    PAIL & 0.42 & 7.2 \\
    OURS & 0.69 & 19.5 \\
    \bottomrule
    \end{tabular}
\end{table}

Table \ref{tab:complexity of model} provides a detailed comparison of the model complexity and inference speed across different reconstruction methods. In terms of model capacity, CTO contains the largest number of parameters, reaching 172M. This relatively high parameter count is mainly attributed to its use of 32 hidden channels in the neural operator architecture, which substantially increases the model width and overall representation capacity. In contrast, our proposed method adopts a more compact configuration with 22 hidden channels, resulting in 19.5M parameters. Although this is larger than lightweight networks such as CvG-Diff (1.85M), MVMS-RCN (0.29M), and PAIL (7.2M), it remains significantly smaller than CTO while still preserving the expressive capability required for accurate CT reconstruction.

Regarding inference efficiency, MVMS-RCN achieves the fastest testing time of 0.05 s per slice due to its lightweight design, followed by CTO with 0.40 s and PAIL with 0.42 s. Our method requires 0.69 s per slice, which is slightly higher than these faster baselines but still more efficient than CvG-Diff (0.89 s). Considering its moderate parameter size and practical inference speed, the proposed method achieves a favorable balance between computational complexity and reconstruction performance. The results indicate that our model avoids the excessive parameter burden of CTO while maintaining clinically feasible reconstruction efficiency.

\section{Discussion}

Under the most challenging unseen 31-view setting, the evaluation metrics exhibit divergent trends: while SONAR and MVMS-RCN achieve nearly identical PSNR values, SONAR yields a slightly lower RMSE, whereas MVMS-RCN obtains a marginally higher SSIM. Given its small magnitude relative to the reported standard deviations, this SSIM difference reflects differing reconstruction biases rather than superior structural recovery. MVMS-RCN utilizes a multiscale geometric correction module involving repeated restriction, prolongation, and error-smoothing operations ~\cite{fan2024mvms}. Under severe angular undersampling, this architecture tends to preferentially preserve large-scale anatomical consistency at the expense of local high-frequency details. This smoothing effect reduces noise and streak artifacts, which can elevate SSIM scores even when subtle structures are suppressed, as SSIM emphasizes local structural consistency over direct high-frequency detail preservation. Visual comparisons corroborate this quantitative trend: MVMS-RCN produces relatively smooth anatomical regions, while SONAR better retains local textures and boundaries. Consequently, the variation across PSNR, RMSE, and SSIM reflects a fundamental trade-off between global structural smoothness and fine-detail preservation, rather than a mathematical contradiction.

The experimental results validate the two central components of SONAR. First, the range–null-space error decomposition demonstrates that SONAR reduces structured errors in both spaces. The diminished range-space error confirms that recovering poorly observable information does not compromise data consistency with the acquired projections. Concurrently, the reduction of anatomically correlated patterns in the null space indicates that the learned complementary-view information robustly constrains image variations unobservable from the measured views alone. Second, the A-UDNO ablation study isolates the contribution of the anisotropic projection-domain operator. While A-UDNO and isotropic UDNO perform similarly under the 47-view training configuration, isotropic UDNO degrades substantially when transferred to 37- and 62-view settings. Because this performance gap emerges primarily outside the training discretization, the advantage of A-UDNO extends beyond increased fitting capacity. By explicitly modeling periodic angular geometry and employing direction-dependent angular and detector supports, A-UDNO significantly enhances the transferability of the learned projection operator across different sampling densities.

The ability to directly re-discretize the model onto finer target grids—without parameter updates or post-hoc interpolation—validates the underlying function-space formulation of SONAR. By learning continuous representations rather than discrete pixel-to-pixel mappings, the architecture natively preserves anatomical boundaries across varying resolutions. Furthermore, the clinical PCCT evaluations confirm that this theoretical capability translates to practical robustness against real-world measurement noise and complex physical variations. Ultimately, these outcomes demonstrate that integrating geometry-aware continuous operators with the explicit separation of physical and learned uncertainties provides a fundamental mechanism for cross-discretization generalization, extending well beyond the specific configurations seen during training.

Despite these advantages, several limitations require further investigation. First, the complementary-view operator used in the practical CT implementation is not an exact orthogonal projector onto $\operatorname{Null}(\mathbf{A})$; rather, it provides coordinates sensitive to both null- and near-null-space variations. Consequently, the quality of the resulting pseudo-measurements relies on the accuracy of the pretrained A-UDNO and may degrade under extremely sparse sampling, irregular angular patterns, or substantial distribution shifts. Second, although the clinical PCCT data inherently contain realistic measurement noise, the current study does not systematically evaluate controlled photon-count levels, varying dose conditions, electronic noise, or geometric model perturbations. Therefore, the independent contributions of noise robustness versus cross-sparsity generalization remain to be quantified. Third, the clinical evaluation is restricted to wrist data acquired from a single PCCT system, utilizing two-dimensional fan-beam data rebinned from helical acquisitions. Since full-view FBP reconstructions serve as reference images, they may introduce residual noise and reconstruction bias. Generalizing these findings to other anatomies, scanner systems, acquisition trajectories, and native three-dimensional reconstructions requires further validation. Finally, the cross-resolution evaluation is currently constrained to grids up to $512\times512$. Future work will scale this to $768\times768$ and $1024\times1024$ grids, accounting for detector sampling, system modulation transfer characteristics, and computational memory limits—recognizing that a larger reconstruction matrix does not inherently guarantee an increase in physically resolvable image information. Future studies will also explore controlled noise experiments, confidence-adaptive weighting of the two residual branches, and computationally efficient cascade implementations.

\section{Conclusion}
This work presented SONAR, a structure-consistent neural operator for
null-space-aware sparse-view CT reconstruction. SONAR represents poorly
observable image information through learned complementary-view
coordinates rather than directly estimating the complete high-dimensional
null-space component. Physical measurement residuals and learned
pseudo-measurement residuals are explicitly separated and independently
regularized according to their image-domain structural effects. An
anisotropic neural operator further accounts for the distinct angular and
detector-domain geometry of CT sinograms, while continuous image-domain
operators enable re-discretization across spatial grids. Experiments on
the simulated AAPM dataset and clinical MARS PCCT data demonstrate
consistent reconstruction performance across seen and unseen angular
sampling densities, different energy bins, and unseen image resolutions.
The error-decomposition and ablation studies further support the
effectiveness of the dual-residual structural-consistency formulation and
the anisotropic operator design. Overall, SONAR provides a physically
interpretable and discretization-flexible framework for accurate
sparse-view CT reconstruction, while offering a foundation for future
extensions to noise-adaptive, cross-geometry, and three-dimensional
imaging.
\bibliographystyle{IEEEtran}
\bibliography{refs}
\end{document}